\documentclass[natbib2009,numberedappendix]{emulateapj}
\usepackage{natbib}
\usepackage{amsmath}
\usepackage{multirow,tabu}
\usepackage{color}

\newcommand{\hii}{\ion{H}{2}}
\newcommand{\cii}{\mbox{[\ion{C}{2}]}}
\newcommand{\oi}{\mbox{[\ion{O}{1}]}}
\newcommand{\nit}{[\ion{N}{2}]}
\newcommand{\nitnp}{\ion{N}{2}}
\newcommand{\siii}{[\ion{S}{3}]}

\newcommand{\nii}{[\ion{N}{2}]~122~$\mu$m}
\newcommand{\niii}{[\ion{N}{2}]~205~$\mu$m}
\newcommand{\ned}{$n_{\rm e}$}
\newcommand \beqa	{\begin{eqnarray}}
\newcommand \eeqa	{\end{eqnarray}}

\newcommand \K{\,{\rm K}}
\newcommand \Ha{{\rm H}}
\newcommand \cm{\,{\rm cm}}
\newcommand \s{\,{\rm s}}
\newcommand \beq{\begin{equation}}
\newcommand \eeq{\end{equation}}
\newcommand \Msol{M_{\odot}}
\newcommand \Lsol{L_{\odot}}
\newcommand \yr{\,{\rm yr}}
\newcommand \erg{\,{\rm erg}}

\shorttitle{The Warm Ionized Medium in Nearby Galaxies}
\shortauthors{Herrera-Camus et al.}

\begin{document}

\title{The Ionized Gas in Nearby Galaxies as Traced by the [\ion{N}{2}]~122 and 205~$\mu$m Transitions}


\author{R.~Herrera-Camus\altaffilmark{1,2}, A.~Bolatto\altaffilmark{1}, J.D.~Smith\altaffilmark{3}, B.~Draine\altaffilmark{4}, E.~Pellegrini\altaffilmark{5}, M.~Wolfire\altaffilmark{1}, K.~Croxall\altaffilmark{6}, I.~de~Looze\altaffilmark{7}, D.~Calzetti\altaffilmark{8}, R.~Kennicutt\altaffilmark{7}, A.~Crocker\altaffilmark{9}, L.~Armus\altaffilmark{10}, P.~van~der~Werf\altaffilmark{11}, K.~Sandstrom\altaffilmark{12}, M.~Galametz\altaffilmark{13}, B.~Brandl\altaffilmark{11,14}, B.~Groves\altaffilmark{15},  D.~Rigopoulou\altaffilmark{16}, F.~Walter\altaffilmark{17}, A.~Leroy\altaffilmark{6}, M.~Boquien\altaffilmark{7}, F.~S.~Tabatabaei\altaffilmark{18},  P.~Beirao\altaffilmark{19}.}

\altaffiltext{1}{Department of Astronomy, University of Maryland, College Park, MD 20742, USA.}
\altaffiltext{2}{Max-Planck-Institut  f\"{u}r extraterrestrische Physik, Giessenbachstr., 85748 Garching, Germany}
\altaffiltext{3}{Department of Physics and Astronomy, University of Toledo, 2801 West Bancroft Street, Toledo, OH 43606, USA}
\altaffiltext{4}{Department of Astrophysical Sciences, Princeton University, Princeton, NJ 08544, USA}
\altaffiltext{5}{Zentrum f\"{u}r Astronomie der Universit\"{a}t Heidelberg, Institut f\"{u}r Theoretische Astrophysik, Albert-Ueberle-Str. 2, 69120 Heidelberg, Germany}
\altaffiltext{6}{Department of Astronomy, The Ohio State University, 4051 McPherson Laboratory, 140 West 18th Avenue, Columbus, OH 43210, USA}
\altaffiltext{7}{Institute of Astronomy, University of Cambridge, Madingley Road, Cambridge, CB3 0HA, UK}
\altaffiltext{8}{Department of Astronomy, University of Massachusetts, Amherst, MA 01003, USA}
\altaffiltext{9}{Department of Physics, Reed College, Portland, OR 97202, USA}
\altaffiltext{10}{Spitzer Science Center, California Institute of Technology, MC 314-6, Pasadena, CA 91125, USA}
\altaffiltext{11}{Leiden Observatory, Leiden University, P.O. Box 9513, 2300 RA Leiden, The Netherlands}
\altaffiltext{12}{Center for Astrophysics and Space Sciences, Department of Physics, University of California, San Diego, 9500 Gilman Drive, La Jolla, CA 92093, USA}
\altaffiltext{13}{European Southern Observatory, Karl Schwarzschild Strasse 2, D-85748 Garching, Germany}
\altaffiltext{14}{Delft University of Technology, Faculty of Aerospace Engineering,
Kluyverweg 1, 2629 HS Delft, The Netherlands}
\altaffiltext{15}{Research School of Astronomy \& Astrophysics, Australian National University, Canberra, ACT 2611, Australia}
\altaffiltext{16}{Department of Physics, University of Oxford, Keble Road, Oxford OX1 3RH, UK}
\altaffiltext{17}{Max-Planck-Institut f\"{u}r Astronomie, K\"{o}nigstuhl 17, D-69117 Heidelberg, Germany}
\altaffiltext{18}{Instituto de Astrof\'isica de Canarias, C/ V\'ia L\'actea, s/n, 38205, San Crist\'obal de La Laguna, Tenerife, Spain}
\altaffiltext{19}{Observatoire de Paris, 61 avenue de l'Observatoire, F-75014 Paris, France}

\begin{abstract}
The \nit~122 and 205 $\mu$m transitions are powerful tracers of the ionized gas in the interstellar medium. By combining data from 21 galaxies selected from the Herschel KINGFISH and Beyond the Peak surveys, we have compiled  141 spatially resolved regions with a typical size of $\sim$1~kiloparsec, with observations of both \nit\ far-infrared lines.  We measure \nit~122/205 line ratios in the $\sim0.6-6$ range, which corresponds to electron gas densities $n_{\rm e}\sim1-300$~cm$^{-3}$, with a median value of $n_{\rm e}=30$~cm$^{-3}$. Variations in the electron density within individual galaxies can be as a high as a factor of $\sim50$, frequently with strong radial gradients. We find that $n_{\rm e}$ increases as a function of infrared color, dust-weighted mean starlight intensity, and star formation rate surface density ($\Sigma_{\rm SFR}$). As the intensity of the \nit\ transitions is related to the ionizing photon flux, we investigate their reliability as tracers of the star formation rate (SFR). We derive relations between the \nit\ emission and SFR in the low-density limit and in the case of a log-normal distribution of densities. The scatter in the correlation between \nit\ surface brightness and $\Sigma_{\rm SFR}$ can be understood as a property of the $n_{\rm e}$ distribution. For regions with $n_{\rm e}$ close to or higher than the \nit\ line critical densities, the low-density limit \nit-based SFR calibration systematically underestimates the SFR since the \nit\ emission is collisionally quenched. Finally, we investigate the relation between \nit\ emission, SFR, and $n_{\rm e}$ by comparing our observations to predictions from the MAPPINGS-III code.
\end{abstract}

\keywords{galaxies: star formation --- galaxies: ISM --- ISM: structure --- infrared: galaxies}

\section{{\bf Introduction}} 

Infrared transitions are a powerful tool for investigating the neutral
and ionized gas in the interstellar medium (ISM). At wavelengths
greater than 100~$\mu$m, the brightest lines in
star-forming galaxies are the \cii~158~$\mu$m and the \nit~122 and
205~$\mu$m fine structure transitions \citep{rhc_wright91,
rhc_bennett94, rhc_malhotra01, rhc_brauher08, rhc_zhao13,rhc_zhao16}. 
While the \cii\ line arise from both the neutral and the ionized
gas, the ionization potential of nitrogen is about $\sim$0.9~eV
higher than that of hydrogen, which implies 
that \nit\ lines originate only from the ionized gas.
This, combined with the fact that far-infrared lines
are affected by dust only in extreme cases, makes the 
\nit~122 and 205~$\mu$m transitions powerful means
to study the properties of the ionized ISM.

The pair of infrared \nit\ lines is the result of the splitting of the
ground-state of N$^{+}$ into three fine-structure levels. These levels
are excited primarily by electron collisions, and the critical
densities ($n_{\rm crit}$) for the resulting \nit~121.89~$\mu$m
($\,^3{\rm P}_2 \rightarrow \,^3{\rm P}_1$) and \nit~205.19~$\mu$m
($\,^3{\rm P}_1 \rightarrow \,^3{\rm P}_0$) transitions are 290~cm$^{-3}$ and
44~cm$^{-3}$, respectively \citep[assuming $T\approx8,000$~K;][]{rhc_hudson04}. 
These excitation conditions imply that
the \niii\ power per N$^+$ scales linearly with electron density ($n_{\rm e}$) up to
$\sim$10~cm$^{-3}$, growing increasingly more slowly with $n_{\rm e}$
until leveling off at $n_{\rm e}\sim60-70$~cm$^{-3}$ \citep[e.g., see
Figure~8 in][]{rhc_langer15}. Above $n_{\rm e}\sim 10 {\rm cm}^{-3}$,
the \nit~122 to 205~$\mu$m line ratio starts to increase from its base value of
$\sim$0.6 \citep{rhc_tayal11} as a function of $n_{\rm e}$, 
until the electron density of the gas approaches the critical density of the \nii\ line 
As we show in Figure~\ref{ne_comparison},
\nit\ based \ned\ measurements of the
photoionized gas have been made for the Galactic plane \citep{rhc_bennett94, rhc_goldsmith15},
(ultra-)luminous infrared galaxies ((U)LIRGs) \citep{rhc_zhao16}, and a handful of other sources: M~82
\citep[$\sim$180~cm$^{-3}$,][]{rhc_petuchowski94}, Carina nebula
\citep[$\sim$28~cm$^{-3}$,][]{rhc_oberst06,rhc_oberst11}, the central
region of NGC~1097 \citep[$\sim$160~cm$^{-3}$,][]{rhc_beirao12}, M~51
\citep[$\sim$8~cm$^{-3}$,][]{rhc_parkin13}, NGC~891
\citep[$\sim$10-100~cm$^{-3}$,][]{rhc_hughes14} and the central region
of IC~342 \citep[$\sim$110~cm$^{-3}$,][]{rhc_rigopoulou13}. 

Another interesting application of the \nit\ far-infrared lines is to
use them as tracers of star formation activity. This use is motivated by
the fact that the \nit\ lines arise from gas ionized by O and
early-B type stars, thus 
providing a direct measurement of the ionizing photon rate, which is
directly related to the star formation rate (SFR)
\citep{rhc_bennett94,rhc_mckee97}. The other advantage is that the
\nit\ far-infrared line emission can be observed in high-redshift
galaxies by ground-based observatories like the Atacama Large
Millimeter Array \citep{rhc_ferkinhoff11,
rhc_combes12,rhc_nagao12,rhc_decarli12,rhc_ferkinhoff15}. Empirical
calibrations of the star formation rate (SFR) based on \nit~122 and
205~$\mu$m luminosities have been derived based on observations of nearby galaxies
\citep[e.g., M83, M51, NGC 891; ][]{rhc_wu15,rhc_hughes16}, and samples of
star-forming and (U)LIRGs by
\cite{rhc_farrah13} and \cite{rhc_zhao13,rhc_zhao16}. These calibrations provide
SFR estimates with an uncertainty of a factor of $\sim3$ for
star-forming galaxies with infrared luminosities below
$\sim10^{11.5}$~L$_{\odot}$; for more luminous galaxies, these
calibrations tend to underestimate the SFR by
factors that can be as high as $\sim10$. On the modeling side,
\cite{rhc_orsi14}, based on the ``Semi-Analytic galaxies'' model \citep[SAG;][]{rhc_cora06,rhc_orsi14}
and the photoionization code MAPPINGS-III \citep{rhc_kewley01, rhc_groves04}, studied the evolution of 
the correlation between the \niii\ luminosity and SFR from redshift $z=5$ to the present. 
They find a \nit-SFR correlation that is consistent,
within the scatter, with the results from \cite{rhc_zhao13}.

One the most important limitations for using the \nit\ emission as a star formation tracer is the decrease in the ratio between the \nit\ lines and the FIR luminosity (which is proportional to the SFR) observed in some local and high redshift galaxies \citep{rhc_fischer10,rhc_gracia-carpio11,rhc_decarli12,rhc_farrah13,rhc_decarli14,rhc_walter09,rhc_zhao13}. This so-called ``\nit-deficit'' may arise from environments with high ionization parameters where ionizing photons from dusty \hii\ regions are intercepted by dust \citep[][]{rhc_luhman03,rhc_gracia-carpio11}, an increased fraction of nitrogen in N$^{++}$ and N$^{+++}$ near very early-type O stars, and the relatively low critical density of the \nit\ far-infrared lines \citep{rhc_langer15}.

Based on a sample of \nit~122 and 205~$\mu$m resolved observations of 21 nearby galaxies by {\it Herschel}, the goal of this paper is twofold. First, to measure the beam-averaged electron density of the low-excitation \hii\ gas and explore any potential dependence with the environment (metallicity,  radiation field strength, etc). Second, to derive \nit-based star formation rate calibrations and study their reliability. This paper is organized as follows. In Section~2 we describe the observations and the sample under investigation. In Section~3 we use the \nit~122 to 205~$\mu$m line ratio to measure electron densities, and explore the properties of the ionized gas. In Section~4 we investigate the connection between \nit\ emission, star formation activity and electron density. Finally, in Section~5 we summarize our main conclusions.

\section{{\bf Sample and methods}} 

Our study focuses on 21 spiral galaxies that are part of the ``Key Insights on Nearby Galaxies: A Far-Infrared Survey with Herschel'' \citep[KINGFISH; P.I.][]{rhc_kennicutt11} and {\it ``Beyond the Peak''} (BtP; P.I. J. D. Smith) samples. Combined, these surveys provide deep photometric and spectroscopic measurements of 20 centrally pointed regions and one extra-nuclear region (NGC~5457).  According to their optical spectral properties, 12 of these 21 galaxies show signatures of active galactic nuclei (AGN) emission \citep{rhc_ho97,rhc_moustakas10}; however, galaxies in the KINGFISH sample have been selected not to have a strong AGN contribution. Our  galaxies cover 
a total infrared luminosity range of $L_\text{TIR} \sim 10^{8.9} - 10^{10.7} L_{\odot}$ \citep{rhc_dale12} and a global metallicity range of $12 + \text{log(O/H)} \sim 8.1 - 8.9$ for the \cite{rhc_pilyugin05} calibration (PT05) and $12 + \text{log(O/H)} \sim 8.7 - 9.2$ for the \cite{rhc_kobulnicky04} calibration (KK04) \citep{rhc_moustakas10}. The SPIRE-FTS beam size at 205~$\mu$m is $16\farcs6$ \citep{rhc_makiwa13};  given the range of distances in our sample ($\sim 3.5-30.6$~Mpc), we cover a range of spatial resolutions that goes from $\sim$0.3~kpc for NGC~2976 to $\sim$2.5~kpc for NGC~1266, with a median value of $\sim0.8$~kpc. For more information on the properties of the KINGFISH and BtP galaxies we refer to Table~1 in \cite{rhc_kennicutt11}.

\begin{figure}
\epsscale{1}
\plotone{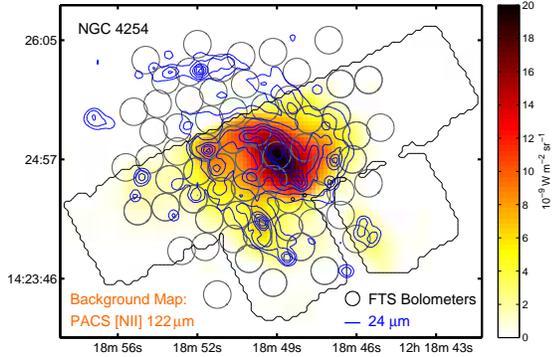}
\caption{PACS \nii\ image of NGC~4254. The color scale shows the surface brightness in units of 10$^{-9}$~W~m$^{-2}$~sr$^{-1}$. The black contours delineate the areas where \nii\ was observed and the blue contours show the 24~$\mu$m dust continuum emission. The grey circles show the distribution of the 17$\arcsec$ SPIRE-FTS bolometers used to detect the \niii\ line. The axes show the RA(J2000) and DEC(J2000) position coordinates. \label{galaxy_map}}
\end{figure}

\begin{figure*}
\epsscale{1}
\plotone{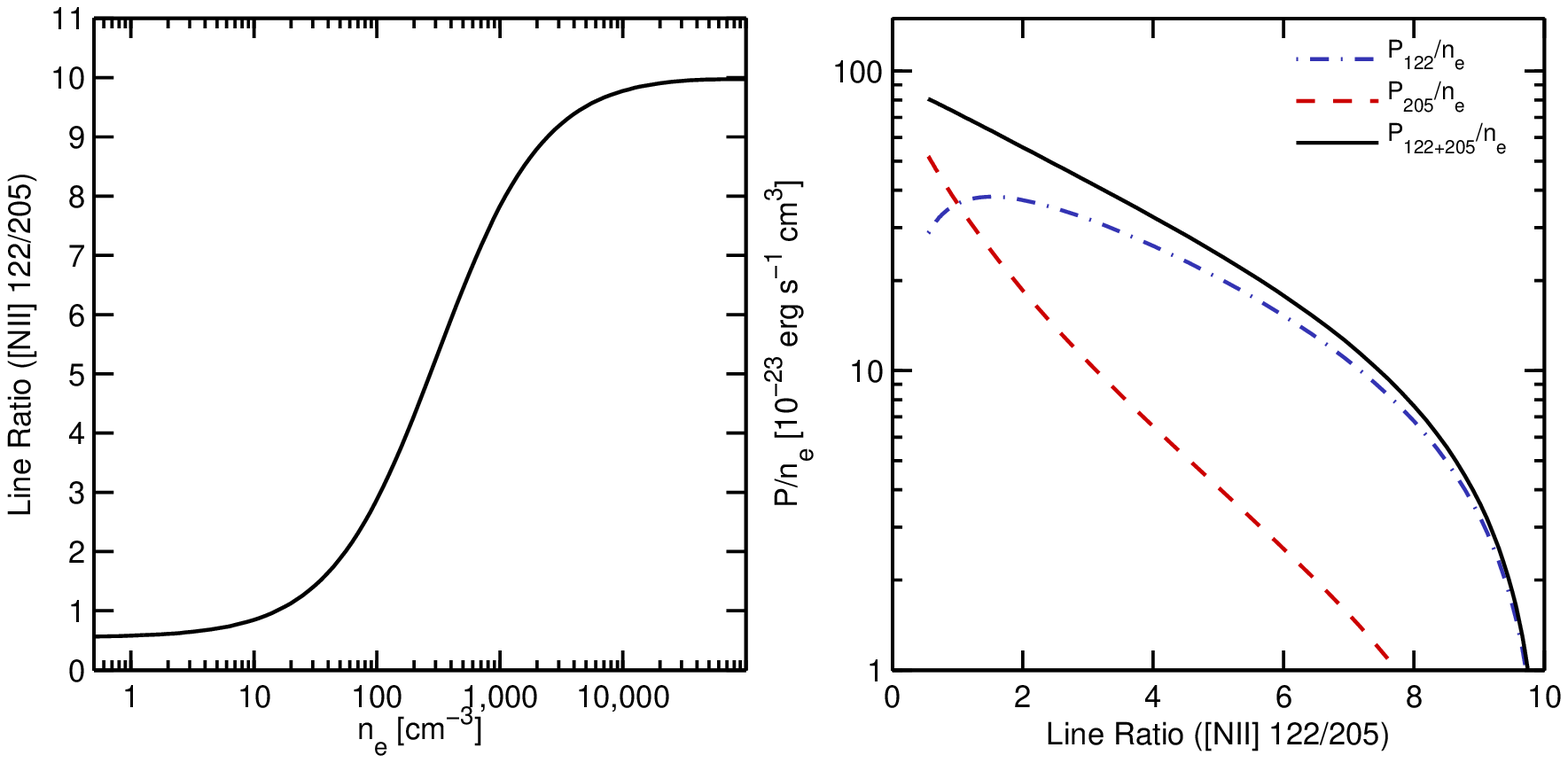}
\caption{{\it (Left)} Ratio of the \nii\ and \niii\ transitions as a function of electron density $n_{e}$. The theoretical curve was calculated using \cite{rhc_tayal11} electron collision strengths and shows how the \nit~122/205 line ratio can be used as a probe of the electron density of the low-excitation, warm ionized gas in the $\sim10-1,000$~cm$^{-3}$ range. We have assumed an electron temperature of $T=8000$~K. {\it (Right)}  $P/n_e$, where $P$ is the power radiated per N+ ion in the \nii\ transition (blue), the \niii\ transition (red), and the sum of both \nit\ transitions (black), as a function of the \nit~122/205 line ratio. At \nit~122/205 line ratios greater than $\sim1$, the total power radiated per ion starts to be dominated by the \nii\ transition. \label{ne_power}}
\end{figure*}

\subsection{Spectroscopic Data}

Observations of the \nii\ transition were carried out with the {\it Herschel} Photodetector Array Camera \& Spectrometer \citep[PACS,][]{rhc_poglitisch10} on board {\it Herschel} as part of the KINGFISH survey. The observations were obtained in the Un-Chopped Mapping mode and reduced using the {\it Herschel} Interactive Processing Environment (HIPE) version~11.0. The reductions applied the standard spectral response functions, flat field corrections, and flagged instrument artifacts and bad pixels \citep[see][]{rhc_poglitisch10, rhc_kennicutt11}. Transient removal was performed using a custom treatment designed for the KINGFISH pipeline. In-flight flux calibrations were applied to the data. We mitigated the spatial undersampling of the beam by using a half pixel dither and then drizzle the maps \citep[details in][]{rhc_kennicutt11}. We then integrated in velocity the \nii\ line from the reduced cubes producing moment zero maps with $2\farcs3$ pixels. The PACS full width half maximum\footnote{\texttt{http://herschel.esac.esa.int/Docs/PACS/pdf/pacs\_om.pdf}} (FWHM) at 122~$\mu$m is $\approx10\arcsec$. The calibration uncertainty on PACS  is  $\sim20$\% \citep{rhc_croxall13}. For a detailed description on the reduction and processing of the KINGFISH FIR spectral maps we refer to \cite{rhc_croxall13}.

The \niii\ transition was observed with the SPIRE Fourier Transform Spectrometer
\citep[SPIRE-FTS,][]{rhc_griffin10} as part of the BtP survey. For a detailed description on the reduction and processing of the spectral maps we refer to \cite{rhc_pellegrini13} and Pellegrini et al. 2016 (in prep.). Hereafter, we will refer to the area covered by a single SPIRE-FTS beam as an ``individual region''. The SPIRE-FTS instrumental spectral resolution is not sufficient to resolve line profiles in BtP. Profile fitting was done with a fixed-width sinc function, where we fit for the line position and peak. Thus the dominant uncertainty in the \niii\ line flux measurement is in uncertainty of the peak above the continuum. To estimate the uncertainty in the continuum under the line we measure the $1-\sigma$ standard deviation in the residual of the fit and take this to be the uncertainty in 1-spectral element. As the noise pattern in the continuum is correlated (fringe pattern) we scale the uncertainty in a single spectral pixel by the number of pixels under an unresolved line to estimate the uncertainty in our fluxes.

The calibration of the SPIRE-FTS spectra depends on the intrinsic source structure. The reduction and calibration routines use gains that are optimized for sources that, compared to the beam size, are either much smaller (point sources) or much more extended. The calibration scale differs by as much as a factor of $\sim2$ between compact and extended \citep[e.g., see Figure~4 in][]{rhc_wu13}.

To address the issue of what is the precise gain that needs to be applied to our \niii\ observations, we compare the SPIRE-FTS synthetic continuum photometry with the SPIRE 250~$\mu$m continuum flux from imaging (the synthetic photometry is derived by applying the 250~$\mu$m SPIRE filter gain curve to the FTS spectrum). The synthetic and continuum photometry need to match, so we attribute any difference between the two to the gain correction due to the intrinsic structure of the source. 

Our SPIRE-FTS data were first reduced assuming a point-source distribution. Based on the method described above, we correct our data by scaling the surface brightness at 205~$\mu$m of each bolometer by the ratio between the synthetic and imaging photometry at 250~$\mu$m. Note that our correction is not based on a constant \niii-to-continuum ratio, but under the assumption that the gain factor that applies to the 250~$\mu$m continuum also applies to the 205~$\mu$m data. The typical 250~$\mu$m flux of our regions is $\sim1.8$~Jy, with an uncertainty less than 10\% of this value. The mean value of this correction is 1.1.

\subsection{Supplementary Data}

The supplementary data available includes: (1) near and mid-infrared (24~$\mu$m) data from the {\it Spitzer} Infrared Nearby Galaxy Survey \citep[SINGS;][]{rhc_kennicutt03};  (2) FIR maps observed with {\it Herschel} PACS (70, 100  and 160~$\mu$m) and SPIRE (250, 350 and 500~$\mu$m) drawn from the photometric KINGFISH sample \citep{rhc_dale12}; (3) H$\alpha$ narrow-band images corrected for Galactic extinction, with foreground stars masked and the optical \nit\ contribution subtracted \citep{rhc_leroy12}. The latter are drawn mainly from the SINGS \citep{rhc_kennicutt03} and Local Volume Legacy survey \citep[LVL,][]{rhc_dale09}. 

As an example of the data used in this study, Figure~\ref{galaxy_map} shows the \nii\ surface brightness map of the spiral galaxy NGC~4254. The blue contours show the 24~$\mu$m dust continuum emission and the grey circles show the spatial distribution of the 68 SPIRE-FTS bolometers used to observe the \niii\ line emission. For this particular galaxy, the fraction of 205~$\mu$m bolometers that overlap with \nii\ data is 45\%. For the entire sample, this fraction is $23$\%. On the other hand, the overlap between the \nit\ spectroscopic data and the complementary photometric data available (e.g. 24~$\mu$m, H$\alpha$, etc) is nearly complete. 

In Section~4 we compare the BtP and KINGFISH data to a sample of local luminous infrared galaxies. This sample consist of  25 ULIRGs observed in \nii\ emission by \cite{rhc_farrah13} and 44 (U)LIRGs observed in \niii\ emission by \cite{rhc_zhao13} as part of the Great Observatories All-sky LIRG Survey \citep[GOALS; ][]{rhc_armus09}. For these samples of LIRGs we measure SFRs based on the total infrared luminosity (TIR; $L(8-1000~\mu m)$) and the calibration by \cite{rhc_murphy11}. 

\subsection{Models}

\subsubsection{Draine \& Li dust model}

For each BtP galaxy we have maps of dust properties based on the Draine \& Li dust model \citep[DL07; ][]{rhc_draine07}. In brief, the DL07 model considers that dust consists of a combination of carbonaceous and amorphous silicate grains whose grain size distribution and normalization is chosen to match the abundance and average extinction in the Milky Way \citep{rhc_weingartner01}. In the model, the dust is heated by a range of radiation fields $U$, including: (1) a diffuse component that is heated by a single radiation field, $U_{\rm min}$; (2) a more intense component, $U_{\rm min}<U<U_{\rm max}$, that heats dust located near luminous stars (e.g., dust in photodissociation regions heated by OB stars). The dust maps we use in this work are similar to the ones presented in \cite{rhc_aniano12}, and were processed by G. Aniano et al. (in prep.) by fitting the DL07 model to the infrared spectral energy distribution in the $3.6-250$~$\mu$m wavelength range. The output of the fit includes the dust mass, the dust-weighted mean starlight intensity, $\langle U \rangle$, and the fraction $f_{\rm PDR}$ of the dust luminosity produced by	photodissociation regions with $U > 100$.

\subsubsection{MAPPINGS-III photoionization code}

In Section~4 we investigate the relationship between the \nit\ emission, the electron density, and the star formation activity based on the predictions by the shock and photoionization code MAPPINGS-III \citep{rhc_kewley01, rhc_groves04}. This code takes synthetic FUV spectra generated by the {\it Starburst99} code \citep{rhc_leitherer99} and produces model HII region spectra integrated over the full ionized volume. The code incorporates a sophisticated treatment of the dust that includes absorption, charging and photoelectric heating \citep{rhc_groves04}. The final spectra consist of a set of emission lines that include the \nit~122 and 205~$\mu$m transitions. In this work we use the pre-computed grids of MAPPINGS-III generated by \cite{rhc_levesque10}. These grids adopt a wide range of parameters, including: (1) star formation history (continuous  or instantaneous burst), (2) age ($0\leq t_{age} \leq5$~Myr), (3) metallicity ($0.05\leq Z/Z_{\odot}\leq2$), (4) ionization parameter $q$, which is the ratio between the incident ionizing photon flux and the gas density ($10^7\leq q \leq 4\times10^8$~cm~s$^{-1}$), and (4) electron density ($n_{e}=10$ or 100~cm$^{-3}$). We use a set of grids that adopt a characteristic ionization parameter for star-forming galaxies of $q=2\times10^7$~cm~s$^{-1}$ \citep{rhc_kewley02}, and for a fixed electron density of $n_{\rm e}=10$ or 100~cm$^{-3}$, allow the metallicity to vary between $Z=Z_{\odot}$ and $Z=2Z_{\odot}$. We also adopt a continuous star formation history model at 5~Myr, which correspond to the age in the {\it Starburst99} model at which there is a balance between the number of O stars being born and dying \citep{rhc_kewley01}. The SFR calibrations used in this work \citep{rhc_calzetti07,rhc_murphy11} are also based on {\it Starburst99} calculations that assume a continuous star formation history model.

\begin{figure*}
\epsscale{1}
\plotone{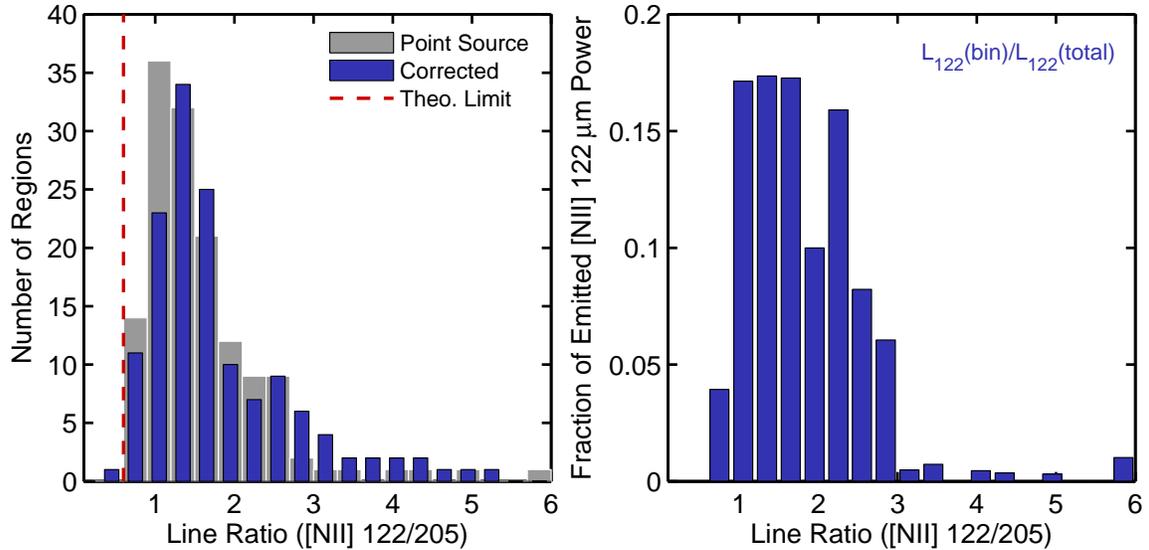}
\caption{(Left) Histograms of the \nit~122/205 line ratio for the 141 regions in our sample when reduced as a point-source (grey) and after applying the extended emission correction (blue). Both distributions are roughly similar, with the main effect of the extended emission correction being a small overall increase of the line ratios. The red dashed line shows the theoretical line ratio of $\approx0.55$ expected for regions of low electron density ($n_{e} \ll n_{\rm crit}$); in this regime the \nit~122/205 line ratio is insensitive to the ionized gas density.  Most of our regions have line ratios above this limit, which means that their line ratios can be used to measure the electron density of the photoionized gas. (Right) Fraction of the emitted \nii\ power per \nit~122/205 line ratio bin. About 40\% of the total \nii\ power arises from regions with \nit~$122/205\gtrsim2$ ($n_{\rm e}\gtrsim50$~cm$^{-3}$).\label{ne_histo}}
\end{figure*}

\subsection{Methods}

In order to assure proper comparison between the supplementary data and the \niii\ observations, we convolved all of our maps to match the SPIRE-FTS FWHM at 205~$\mu$m using convolution kernels constructed using the methodology of \cite{rhc_aniano11}. We then extracted continuum and line fluxes from regions corresponding to the position and sizes of the SPIRE-FTS \niii\ data.

In Section~5 we study the reliability of the \nit\ far-infrared lines as star formation tracers. For this purpose, we measure star formation rate surface densities ($\Sigma_{\rm SFR}$) and SFRs based on a combination of the convolved 24~$\mu$m and H$\alpha$ data following the calibration by \cite{rhc_calzetti07}. This calibration adopts an IMF with $dN/dM \propto M^{-\alpha}$, with $\alpha=-1.3$ in the range $0.1-0.5~{\rm M}_{\odot}$, and $-2.3$ in the range $0.5-120~{\rm M}_{\odot}$. This choice of IMF produces SFRs that, for the same number of ionizing photons, are $\sim14$\% higher than if we change the upper-mass cutoff to 100~M$_{\odot}$, and a factor 1.59 lower if we assume a Salpeter IMF in the range $0.1-100~{\rm M}_{\odot}$. 

We also measure total infrared (TIR) luminosities based on the calibration by \cite{rhc_galametz13} and using the 24, 70, 100 and 160~$\mu$m convolved data.

For the global metallicities of 17 of the 21 galaxies, we use the average between the KK04 and PT05 ``characteristic'' metallicities listed in Table~9 of \cite{rhc_moustakas10}. For the four remaining galaxies with metallicity measurements not available in \cite{rhc_moustakas10}, we use oxygen abundances derived from the Luminosity-Metallicity relation listed in \cite{rhc_kennicutt11}. Nine of the BtP galaxies have measured metallicity gradients \citep{rhc_moustakas10}. However, the regions in these systems have galactocentric distances $\lesssim0.3~R_{25}$ \citep[where $R_{25}$ is the radius of the major axis at the $\mu_{B}=25$~mag~arcsec$^{-2}$ isophote;][]{rhc_deva91,rhc_moustakas10}, therefore the effect of the metallicity gradient is small, and we use the central metallicity.

We correct all surface brightnesses and SFR surface densities for inclination by multiplying $cos(i)$. Inclinations were drawn from the compiled lists in \cite{rhc_hunt15}. We remove NGC~4631 from the analysis due to its high inclination \citep[$i\approx86^{\circ}$;][]{rhc_mmateos09}.

\section{{\bf \nit-based electron densities and their dependence on environment}} 

\subsection{Estimating ionized gas densities from the \nit\ fine-structure transitions}

In this section we discuss how the
electron density can be derived from the ratio between the \nii\ and
\niii\ transitions (from now on \nit~122/205 line ratio).
Let $f_i(n_e)$ be the fraction of ${\rm N^+}$ in level $i$, where $i=0$ is
the ground state. The power radiated in fine structure lines is

\beqa
L_\lambda&=&\int n({\rm N^+}) P_\lambda(n_e) dV 
\\
&=& \left[\frac{\rm N^+}{\rm H^+}\right] \int n_e P_\lambda(n_e) dV \label{eq:Llambda}
\\
P_{205\mu{\rm m}}(n_e) &=& f_1(n_e) A_{10} h\nu_{10}
\\
P_{122\mu{\rm m}}(n_e) &=& f_2(n_e) A_{21}h\nu_{21}
~~~.
\eeqa

\noindent where $A_{10}$ and $A_{21}$ are the Einstein coefficients 
for the $1\rightarrow0$ ($205~\mu$m) and $2\rightarrow1$ ($122~\mu$m)
transitions of ${\rm N^+}$, respectively. We have calculated $f_i(n_e)$
for ${\rm N^+}$ levels $i=0,..,4$, using electron collision strengths
from \cite{rhc_tayal11}, and radiative decay rates from
\cite{rhc_galavis97} and \citet{rhc_storey00}, for an assumed electron
temperature $T=8000\K$, and a range of $n_e$. The left panel of
Figure~\ref{ne_power} shows the variation of the \nit~122/205 ratio
with electron density, $n_{\rm e}$. It can be seen that the line ratio
is sensitive to the density of the photoionized gas in the $n_{\rm
e}\sim10-1,000$~cm$^{-3}$ range.

The right panel of Figure~\ref{ne_power} shows $P/n_{\rm e}$, 
where P is the power radiated per N$^{+}$ ion, for each individual \nit\
transition and for their sum, as a function of the \nit~122/205
line ratio. This figure illustrates the effect of collisional quenching
on the \nit\ transitions once the electron density of the gas exceeds
the critical density of the line. For example, 
as the electron density exceeds the critical density
of the \nii\ line --which happens around a line ratio
\nit~$122/205\approx5$ ($n_{\rm e}\approx250$~cm$^{-3}$)-- 
the power radiated per ion by the
\nii\ line starts to decrease at a rate comparable to that of the \niii\ line.


\subsection{Distribution of \nit~122/205 line ratios and electron densities in the Beyond the Peak sample}

\begin{figure*}
\epsscale{1}
\plotone{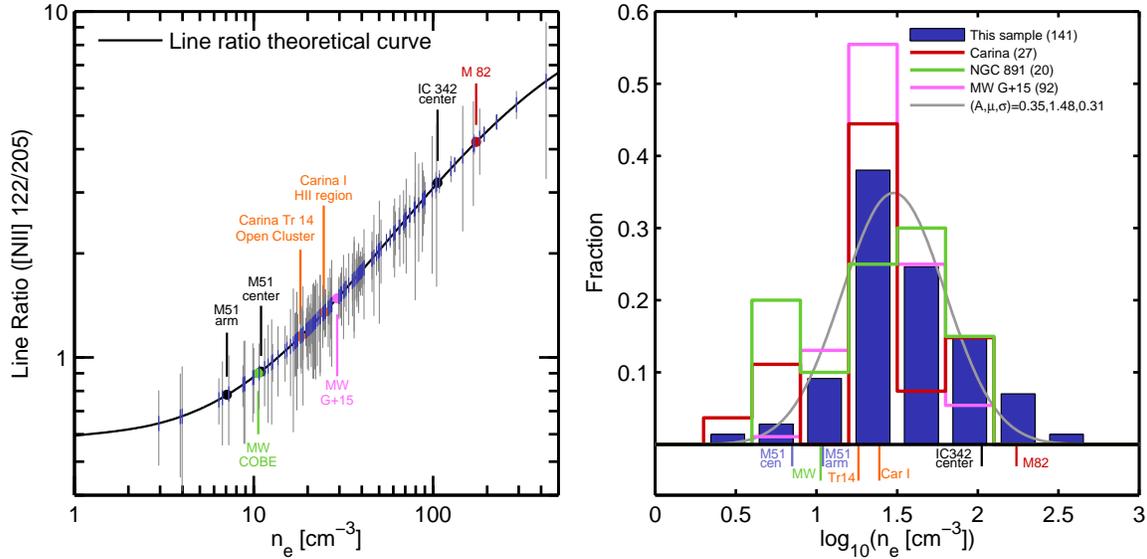}
\caption{(Left) \nit~122/205 line ratio as a function of electron density ($n_{\rm e}$). The blue vertical lines show the observed ratios, and corresponding $n_{\rm e}$ measurements, for 141 regions selected from 21 galaxies in the BtP Sample. The grey vertical lines show the line ratio error. We measure electron densities in the range $n_{\rm e}\sim1-300$~cm$^{-3}$. We also include \nit~122/205 line ratio measurements from M~51 central region and spiral arm \citep{rhc_parkin13}, the Milky Way \citep{rhc_bennett94,rhc_goldsmith15}, the young open cluster Trumpler~14 and the \hii\ region Carina \textrm{I} in the Carina nebula \citep{rhc_oberst11}, IC~342 central region \citep{rhc_rigopoulou13} and M~82 \citep{rhc_petuchowski94}. The median electron density in our sample is $n_{\rm e}\approx30$~cm$^{-3}$, similar to the ionized gas density of the Carina~I \hii\ region. (Right) Distribution of electron densities for the BtP sample (blue bars), 22 regions in the Carina nebula \citep[red,][]{rhc_oberst11}, 92 regions in the Milky Way \citep[G+15;][]{rhc_goldsmith15}, and 20 regions in the edge-on galaxy NGC~891 \citep{rhc_hughes14}. On the bottom of the plot we also show the individual measurements displayed in the left panel. The grey curve shows a Gaussian fit to the BtP distribution of electron densities, and in the legend we list the amplitude (A), mean electron density ($\mu$) and standard deviation ($\sigma$) of the fit.\label{ne_comparison}}
\end{figure*}

\begin{figure}
\epsscale{1.1}
\plotone{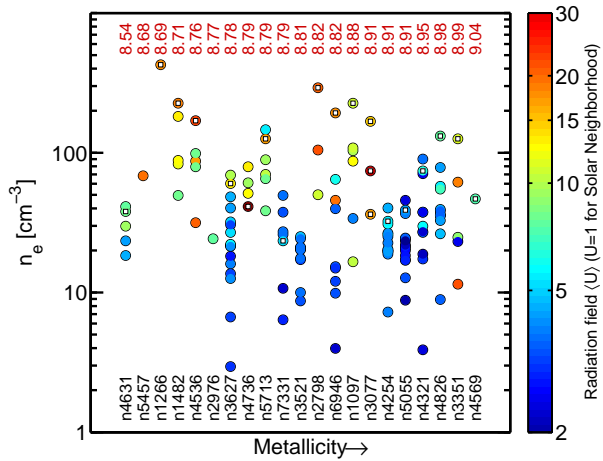}
\caption{Electron density as a function of metallicity for all regions in our sample sorted by galaxy. The name of the galaxy is listed in the bottom part of the plot and the characteristic global value of the Oxygen abundance $12+log(\rm{O/H})$ --measured by \cite{rhc_moustakas10}-- is listed on the top part of the plot in red. Metallicity increases to the right. The color scale corresponds to the dust-weighted mean starlight intensity $\langle U \rangle$ derived from the \cite{rhc_draine07} model. The symbols with inner white squares show the regions where the position of the bolometer is located within a deprojected distance of 400~pc to the center of the galaxy. Within galaxies, and despite the limited spatial coverage of the disk, we observe variations in electron density greater than a factor of $\sim10$ (e.g., NGC~3627, NGC~4826, NGC~6946). For most galaxies, the highest electron density measurements are found in the central regions. Finally, the color scale reveals a clear trend of increasing electron density with radiation field strength. \label{ne_single}}
\end{figure}

For 141 individual regions (defined by the area covered by a single SPIRE-FTS beam at 205~$\mu$m) for which we have observations of both \nit\ transitions with $S/N\geq3$, Figure~\ref{ne_histo} shows the histogram of the observed \nit~122/205 line ratios (left panel) and the fraction of the total emitted \nii\ power per \nit~122/205 bin (right panel). On the left panel, the two \nit~122/205 line ratio distributions show the data calibrated as a point-source (grey) and after applying the extended emission correction (blue). Both distributions are roughly similar, but the corrected version of the data tends to show higher line ratios. This is the direct result of applying the extended emission correction, which accounts for the overestimation of the \niii\ intensity when extracted as a point source (see Section~2.1 for details). From the rest of the paper, we  base all of the results on the extended emission corrected version of the data. The red dashed line shows the line ratio limit of \nit~$122/205\approx0.55$ expected for regions with $n_{e} \ll n_{\rm crit}$ assuming \cite{rhc_tayal11} electron collision strengths (this ratio limit is 0.66 if we assume \cite{rhc_hudson05} electron collision strengths instead). In our sample, there is only one region with a \nit~122/205 line ratio lower than 0.55. This region is located in NGC~4254 and has a \nit~122/205 line ratio of $0.38\pm0.15$. As a cautionary note, this region is one of the few where the extended emission correction increased the \niii\ intensity by a factor of $\sim2$. Therefore, the origin of this low \nit\ line ratio may not be physical. About 40\% of the emitted \nii\ power arises from regions with \nit~$122/205\gtrsim2$, or equivalently, electron densities higher than the critical density of the \niii\ line.

The left panel of Figure~\ref{ne_comparison} shows the theoretical dependence of the electron density on the \nit~122/205 line ratio calculated in Section~3.1 and shown in the left panel of Figure~\ref{ne_power}. 
The observed \nit~122/205 line ratios measured in our 141 regions from the BtP sample are shown as blue vertical lines on top of the curve. The line ratios range between $\sim0.6$ and 6, which corresponds to electron densities in the $\sim3$ to 300~cm$^{-3}$ range. Typical uncertainties in the electron density measurements are of the order of 20\%. Figure~\ref{ne_comparison} also includes \nit~122/205 line ratios observed in the Milky Way with the Cosmic Background Explorer \citep{rhc_bennett94}, M~82 with the Kuiper Airborne Observatory \citep{rhc_petuchowski94}, the young open cluster Trumpler~14 and the \hii\ region Carina \textrm{I} in the Carina nebula observed with the South Pole Imaging Fabry-Perot Interferometer at the Antarctic Submillimeter Telescope and Remote Observatory and the Infrared Space Observatory \citep{rhc_oberst11}, M~51 central and arm regions observed with {\it Herschel} PACS \citep{rhc_parkin13} and the central region of IC~342 observed with {\it Herschel} PACS and SPIRE \citep{rhc_rigopoulou13}. We measure a mean electron density in the BtP sample of $n_{\rm e}\approx30$~cm$^{-3}$, which is similar to the electron density in the young open cluster Trumpler~14, the Carina~\textrm{I} \hii\ region, the average value found in the Galactic plane by \cite{rhc_goldsmith15}, and the median value of 22~cm$^{-3}$ measured in a sample of 12 (U)LIRGs \citep{rhc_zhao16}.

The right panel of Figure~\ref{ne_comparison} shows the resulting distribution of electron densities for the BtP regions. We also plot the literature measurements included in the left panel and we add the distributions of electron densities measured in 20 regions of the edge-on galaxy NGC~891 \citep{rhc_hughes14}, 92 positions in the Galactic plane \citep{rhc_goldsmith15}\footnote{We only included line of sight positions from Table~2 in \cite{rhc_goldsmith15} that have $\geq3\sigma$ line detections in both \nit~122 and 205~$\mu$m transitions.}, and 27 regions in the Carina nebula \citep{rhc_oberst11}. In the latter, the highest density regions are associated with the outskirts of the \hii\ regions Carina~\textrm{I} and \textrm{II}, and the lowest density regions correspond to an extended component detectable all over the $\sim$30~pc map \citep{rhc_oberst06, rhc_oberst11}. The comparison between the BtP data and the external samples reveal the wide range of electron densities present in our sample. On one hand, we are sensitive to a more extended, low-density ionized gas component, like the one that fills the medium in between \hii\ regions in the Carina nebula. On the other hand, we have regions with high electron densities (measured in the central regions of NGC~1097, NGC~4536 and NGC~6946) that are comparable to the ones measured in very active star-forming galaxies, like M~82 and the central region of IC~342.

\subsection{Electron density variations within individual galaxies}

The photoionized gas traced by the \nit\ in our sample spans two orders of magnitude in electron density. In this section we explore the dependence between the electron density and a number of ISM properties that might play a role in these variations, such as metallicity, radiation field strength and star formation activity.

\begin{figure*}
\epsscale{0.85}
\plotone{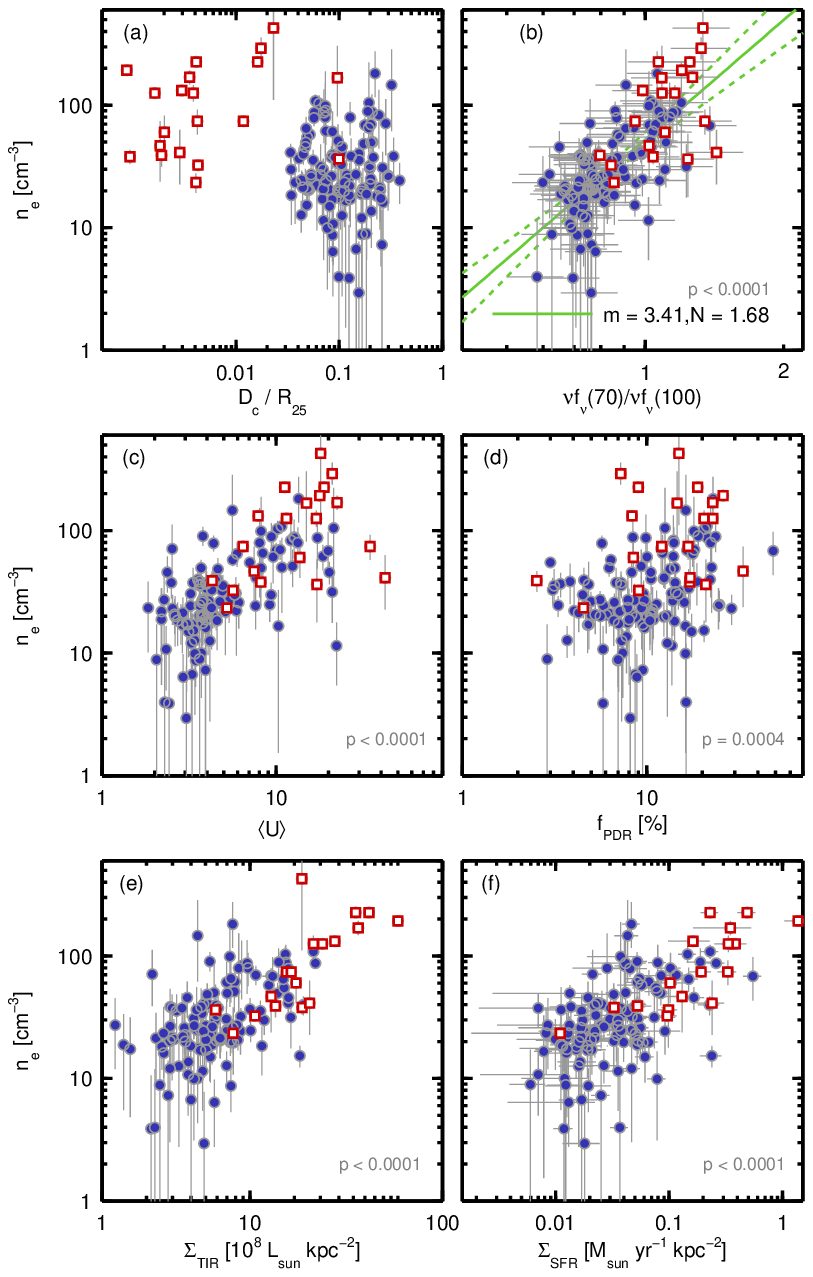}
\caption{Electron density ($n_{\rm e}$) as a function of six parameters. 
The p-values for testing the null hypothesis are shown in the lower right corner of each panel. (a) galactocentric distance (${\rm D_{c}}$) normalized to the radius of the major axis at the $\mu_{B}=25$~mag~arcsec$^{-2}$ isophote \citep[$R_{25}$;][]{rhc_deva91,rhc_moustakas10}. Central regions with deprojected distances less than 400~pc are shown as open red squares. These regions have an average electron density a factor of $\sim3$ higher than the rest of the regions. (b) infrared color ($\nu f_{\nu}(70~\mu{\rm m})/\nu f_{\nu}(100~\mu{\rm m})$). The correlation is good, with a Spearman correlation coefficient of $r=0.73$. The solid and dashed green lines show the best power-law fit to the data (Eq. 5) and the associated confidence bands, respectively. The slope ($m$) and normalization of the fit ($N$) are listed in the bottom-right corner and in Equation~1. (c) dust-weighted mean starlight intensity ($\langle U \rangle$) derived from the \cite{rhc_draine07} model. (d) fraction of the dust luminosity produced by photodissociation regions with $U>100$ ($f_{\rm PDR}$) also derived from the \cite{rhc_draine07} model. (e) total infrared luminosity surface density, $\Sigma_{\rm TIR}$. (f) star formation rate surface density ($\Sigma_{\rm SFR}$) measured using a combination of $24~\mu$m and H$\alpha$ emission following \cite{rhc_calzetti07} calibration. \label{ne_versus}}
\end{figure*}

How much does the electron density vary within the area sampled in each galaxy? To answer this, Figure~\ref{ne_single} shows the electron density of our 141 regions grouped by galaxy. The names of the galaxies are listed in the bottom of the panel, and their characteristic metallicities are listed in the top part of the plot. Galaxies are sorted from left to right in order of increasing metallicity. We use color to include information about the local radiation field strength, derived using the dust-weighted mean starlight intensity $\langle U \rangle$ from the \cite{rhc_draine07} model. Finally, we mark using open circles regions with deprojected distances to the center of the galaxy less than the median physical radius of the SPIRE-FTS bolometer in our sample, i.e. $\approx400$~pc. We measure the distance to the center by calculating the angular separation between the position of the center of the galaxy (taken from the NASA Extragalactic Database) and the position of the SPIRE-FTS bolometers. We then convert the angular separation into a deprojected distance, $D_{\rm c}$, using distances drawn from the compiled list in \cite{rhc_kennicutt11}. From Figure~\ref{ne_single} we notice that the highest electron density measurements within a galaxy tend to be found in its central region (in 16 of the 18 galaxies with central measurements available, the central region corresponds to the highest or the second highest electron density measured in the disk).

Figure~\ref{ne_single} also shows that we are able to detect variations in the electron density within individual galaxies as high as a factor of $\sim50$. One good example is NGC~3627, a Leo Triplet interacting spiral galaxy with a strong bar \citep{rhc_reagan02}. The highest gas density in the system, $n_{\rm e}=71$~cm$^{-3}$, is measured at one of the ends of the bar. This region is characterized by strong star formation activity and high average radiation field strength ($\langle U\rangle \approx10$). The second highest ionized gas density measurement comes from the central region of the galaxy ($n_{\rm e}=40$~cm$^{-3}$, $\langle U \rangle \approx13$), and the lowest ionized gas density ($n_{\rm e}=4$~cm$^{-3}$) is measured in a more quiescent region ($\langle U \rangle \approx3$) located in between the bar and one of the spiral arms. This trend of increasing electron density with radiation field strength is not exclusive to NGC~3627. In fact, Figure~\ref{ne_single} reveals a similar trend for the rest of the sample: while regions with $n_{e}\lesssim20$~cm$^{-3}$ tend to have radiation field strengths of only a few times the radiation field in the solar neighborhood, regions with $n_{e}\gtrsim100$~cm$^{-3}$ tend to have $\langle U \rangle \gtrsim15$. Among nearby galaxies that are not part of the BtP sample, spatial variations of the \niii\ emission are also observed for M~83, NGC~891, M~51 and NGC~4038/9 \citep{rhc_wu15,rhc_hughes16}.

There is not a clear trend of varying electron density with oxygen abundance. However, low metallicity galaxies in our sample tend to have, on average, higher radiation fields and electron densities than the rest of the sample. This could be an observational bias as the spatial coverage of the \nit\ line emission in these low metallicity environments --especially in the case of the \nii\ transition-- is mostly limited to bright, star-forming regions. 

\subsection{Relationship between electron density and the ISM environment}

One of our main goals is to understand the variations in the ionized gas density as a function of the ISM environment. This is possible thanks to the spatial coverage of our target galaxies provided by SPIRE-FTS and the rich characterization of the ISM properties derived from the ancillary data. We start our analysis by studying how the electron gas density changes with radial distance. Panel (a) in Figure~\ref{ne_versus} shows the electron density of the 141 regions in our sample as a function of the deprojected distance to the center of the galaxy normalized by the radius of the major axis at the $\mu_{B}=25$~mag~arcsec$^{-2}$ isophote \citep[$R_{25}$;][]{rhc_deva91,rhc_moustakas10}. The resulting 19 regions with deprojected distances to the center less than 400~pc are shown as red open squares. As we already discussed for Figure~\ref{ne_single}, central regions in galaxies tend to have higher electron densities (median $n_{\rm e}\approx77$~cm$^{-3}$) than those located in the disk (median $n_{\rm e}\approx25$~cm$^{-3}$). 

Panel (b) shows the ionized gas density as a function of the infrared color $\nu f_{\nu}(70~\mu{\rm m})/\nu f_{\nu}(100~\mu{\rm m})$ measured using the PACS 70 and 100~$\mu$m data. There is a good correlation of increasing gas density with increasing $\nu f_{\nu}(70~\mu{\rm m})/\nu f_{\nu}(100~\mu{\rm m})$. The best power-law fit, shown as a solid line in the second panel, yields:

\begin{multline} \label{eq:correlation_uncorr}
{\rm log_{10}(n_{e}/cm^{-3})} = 3.41 \times {\rm log_{10}(\nu f_{\nu}(70)/\nu f_{\nu}(100))}+1.68.
\end{multline}

\noindent The standard deviation around the fit is 0.42~dex. This parametrization could be useful for studies that require an electron density in order to predict the  \niii\ flux based on the \nii\ flux (or vice versa) \cite[e.g.,][]{rhc_zhao13}.

Panels (c) and (d) in Figure~\ref{ne_versus} show the electron density as a function of the dust-weighted mean starlight intensity, $\langle U \rangle$, and the fraction of the dust luminosity produced by photodissociation regions with $U>100$, $f_{\rm PDR}$ (both parameters derived from the \citealt{rhc_draine07} model). The correlation of electron density with starlight intensity is as good as the correlation with infrared color, which is expected given the good correspondence between infrared color and $\langle U \rangle$ (e.g., see Figure~19 in \citealt{rhc_mmateos09}). On the other hand, there is no strong dependence between electron density and $f_{\rm PDR}$ (Spearman correlation coefficient $r=0.38$). 
Finally, panels (e) and (f) show the correlation between electron density and total infrared luminosity surface density, $\Sigma_{\rm TIR}$, and star formation surface density, $\Sigma_{\rm SFR}$. Similar to what is found for the dependences with infrared color or radiation field strength, there is a clear trend of increasing electron density with increasing infrared surface brightness and star formation activity. 

The observed relationship between electron density and star formation activity, or radiation field strength, could have at least two origins. The first one is related to the first stages of the evolution of \hii\ regions. For recent episodes of star formation activity, we expect massive stars to produce very intense radiation fields. The corresponding young \hii\ regions created around these newly-formed stars are more compact than evolved \hii\ regions, characterized by electron densities higher than $n_{e}\approx10^{3}$~cm$^{-3}$, and located in the high-pressure, inner regions of molecular clouds \citep{rhc_franco00}. Although our sensitivity to high electron densities is limited by the critical density of the \nii\ line, we believe that regions in our sample that exhibit high radiation fields and high electron densities could be related to young, compact \hii\ regions. The second possible explanation is based on a thermal pressure argument. Once a young, compact \hii\ region has expanded and reaches pressure equilibrium with its surrounding medium, a high electron density implies a high neutral density of the cold ISM gas in which the \hii\ region is embedded. This high density molecular gas environment provides the conditions for further star formation to occur, establishing a link between the high density of the ionized gas and more intense star formation activity. 

\section{\bf Estimating star formation rates from the \nit\ fine-structure transitions}

In this section we analyze the reliability of the far-infrared \nit\ transitions as tracers of star formation activity. One of the advantages of these lines is that, unlike H$\alpha$ or other optical tracers, they are insensitive to dust extinction and can provide a robust estimate of the ionizing photon rate, $Q_0$. We start by deriving a theoretical relation between \nit\ emission and SFR. This relation depends, among other things, on the nitrogen abundance, the ionization fraction of nitrogen, and the density of the ionized gas. For the latter we study two cases: (1) low-density gas, and (2) gas that follows a log-normal distribution of densities. Finally, we compare our theoretical predictions and the results from the MAPPINGS-III code to the BtP galaxies and other samples of extragalactic objects.

\subsection{Low Density Limit}

To understand how \nit\ emission works as a measure of star formation
we can assume that the excitation is dominated by collisions from the
ground level and balanced by radiative de-excitation. This will be
correct at densities lower than the critical density of the line,
where collisional de-excitation does not
play a role. Thus, we can approximate the power radiated in the
 \nit~122 or 205~$\mu$m lines in a given volume $V$ as

\beq \label{eq:L_nii_low}
L_{\lambda}\cong n_e n({\rm N^+}) q_{\lambda} h \nu V.
\eeq

\noindent 

Here $q_{122} = q_{02}$ and $q_{205} = q_{01}+q_{02}$, where
$q_{01}$ and $q_{02}$ are the collisional excitation coefficients from 
the ground level to level 1 and 2, respectively. $n({\rm N^+})$ is the
density of the ionized nitrogen and we assume that most N$^+$ is in the 
ground level. For an ionization bounded \ion{H}{2} region the 
ionization-recombination balance dictates that

\beq \label{eq:Q0_low}
Q_0 = n_e n({\rm H^+}) \alpha_B V,
\eeq

\noindent 
where the total rate of H photoionizations, $Q_0$, is equal to the
rate of radiative recombinations determined by the case B
recombination coefficient $\alpha_B=3.04\times10^{-13}\cm^3\s^{-1}$ at
$T=8000\K$, and $n(\Ha^+)$ is the number density of ionized hydrogen
atoms. Under these hypotheses, and combining Equations~(\ref{eq:L_nii_low})
and (\ref{eq:Q0_low}), the ionizing photon rate is
proportional to the \nit\ luminosity through (c.f., Eq. 15 in
\citealt{rhc_mckee97})

\beq \label{eq:Q_low}
Q_0 \cong \frac{L_{\lambda} \alpha_B}{q_{\lambda} h \nu_{\lambda}} \frac{n({\rm H^+})}{n({\rm N^+})}.
\eeq

\noindent 
Given the similarities in the ionization potentials (13.6 eV
vs. 14.5 eV), and assuming that 
the nitrogen is only singly ionized (i.e., ${\rm N}^{+}/{\rm N}=1$), with a negligible fraction of higher ionization states (which require photons more energetic than 29.6~eV), the last factor is equal to the
inverse of the gas-phase abundance of nitrogen. 

The nitrogen ionization balance depends on the spectrum of the stellar radiation and the ionization parameter $U_{\rm ion}$, which is defined as the ratio of the ionizing photon density to the hydrogen density $n_{\rm H}$.  Because the N$^{+}$ $\rightarrow$ N$^{++}$ ionization potential is above that for He $\rightarrow$ He$^{+}$, then N$^{++}$ and N$^{+++}$ will be present only where the He is ionized.  Only the hottest O stars (earlier than O8) can ionize He throughout the \hii\ region.  If the He is ionized, the nitrogen will be mainly N$^{++}$ and N$^{+++}$ only if the local ionization parameter $U_{\rm ion} > 10^{-3.1}$ \cite[see Fig. 8 of][]{rhc_abel09}. The median $U_{\rm ion}$ in an \hii\ region is $> 10^{-3}$ for $n_eQ_{0} > 10^{50}{\rm ~cm^{-3}~s^{-1}}$ \cite[see Fig. 3 of][]{rhc_draine11}. Thus in very high density \hii\ regions, or giant \hii\ regions ionized by a cluster of O stars, the nitrogen may be preferentially in N$^{++}$, but in low density \hii\ regions around single stars we expect N to be primarily N$^{+}$.

Assuming solar abundance (N/H)$_\odot=7.4\times10^{-5}$ \citep{rhc_asplund09} and $q_{205} = q_{01}+q_{02}=6.79\times10^{-8}$~cm$^{3}$~s$^{-1}$ \citep{rhc_hudson05}, we measure a median \niii-based global ionizing photon rate for the BtP galaxies of $Q_{0}=1.98\times10^{52}$~photons~s$^{-1}$. Normalized to the covered area, this corresponds to $1.96\times10^{51}$~photons~s$^{-1}$~kpc$^{-2}$. If we use the collisional excitation coefficients from \cite{rhc_tayal11} instead, then $q_{01}+q_{02}=5.1\times10^{-8}$~cm$^{3}$~s$^{-1}$, which increases the ionizing photons rates by a factor of 1.3. The median ionizing photon rate surface density in our sample is about 3 times the ionizing photon rate measured by \cite{rhc_bennett94} inside the solar circle in our Galaxy. Using the \nii\ transition --in principle a better tracer than \niii\ due to its higher critical density-- yields an average ionizing photon rate surface density a factor of $\sim4$ higher than the one based on the \niii\ emission. 

To relate the ionizing photon rate $Q_{0}$ to the SFR, we use the fact that for 
the default {\it Starburst99} IMF with an upper-mass cutoff of 120~M$_{\odot}$ (see Section~2.4 for details), and steady star formation for $10^8$~yr, the star formation rate and rate of
production of photoionizing photons are related by

\beq \label{eq:Q2SFR}
\frac{Q_0}{f_{\rm ion}} = 1.60\times10^{53}\s^{-1} \frac{SFR}{{\rm \Msol\yr^{-1}}},
\eeq

\noindent 
where $f_{\rm ion}$ is the fraction of H ionizing photons emitted by
stars that photoionize H or He; i.e., $(1-f_{\rm ion})$ is the
fraction absorbed by dust. According to the discussion of dusty
\ion{H}{2} regions by \cite{rhc_draine11}, we expect $f_{\rm
ion}\approx1$. Also note that if we change the IMF upper-mass cutoff
to 100~M$_{\odot}$ \citep[e.g.,][]{rhc_murphy11}, the rate of
ionizing photons per SFR decreases by $\sim14\%$.

Combining Equations~(\ref{eq:Q_low}) and (\ref{eq:Q2SFR}) we find that
the SFR as a function of the \nit\ luminosity in the low density limit is
given by:

\beq \label{eq:SFR_low1}
\begin{split}
\frac{\rm SFR}{{\rm \Msol~\yr^{-1}}} = 1.49\times10^{-7}
\left(\frac{6.79\times10^{-8}~{\rm cm}^{3}~{\rm s}^{-1}}{q_{01}+q_{02}}\right) \\
\times \left(\frac{({\rm N/H})_\odot}{\rm N^{+}/H^{+}}\right)
\frac{L_{205}}{\Lsol}
~~~,
\end{split}
\eeq

and 

\beq \label{eq:SFR_low2}
\begin{split}
\frac{\rm SFR}{{\rm \Msol~\yr^{-1}}} = 2.35\times10^{-7}
\left(\frac{2.57\times10^{-8}~{\rm cm}^{3}~{\rm s}^{-1}}{q_{02}}\right) \\
\times \left(\frac{({\rm N/H})_\odot}{\rm N^{+}/H^{+}}\right)
\frac{L_{122}}{\Lsol}
~~~.
\end{split}
\eeq

We can see through this example that a measurement of SFR using the
fine structure lines of \nit\ is possible, but it will depend on the
abundance of nitrogen, its ionization state, and ultimately the
density of the ionized region, as collisional de-excitation will be
important at densities of interest.

\subsection{Effect of a Distribution of Densities}

The photoionized gas in a galaxy will generally have a wide range of
electron densities, from compact \hii\ regions to diffuse photoionized
gas. The balance between photoionization and radiative recombination can
then be expressed as 

\beq
Q_0 = \int \alpha_B n(\Ha^+)n_e dV
= \int \alpha_B n(\Ha^+)n_e \frac{dV}{d\ln n_e} d\ln n_e .
\eeq

In order to assess the effect on $Q_0$ of having a range of electron densities, we parametrize the distribution of electron densities using a log-normal distribution

\beq
\alpha_B n(\Ha^+)n_e \frac{dV}{d\ln n_e} =
\frac{Q_0}{\sqrt{2\pi\sigma^2}}
\exp\left[-\frac{(\ln(n_e/n_{e0}))^2}{2\sigma^2}\right] \label{eq:ne_dist}
~~~;
\eeq

\noindent 
$n_{e0}$ is then a characteristic electron density for the recombining
gas, while $\sigma$ represents the width of the distribution of
electron densities. The case $\sigma=0$ correspond to uniform
density. For $\sigma=1$, for example, if the characteristic
electron density is $n_{\rm e0}=100$~cm$^{-3}$, then the 1-$\sigma$
(68\%) and 2-$\sigma$ (95.5\%) confidence intervals encompass the
density ranges $n_{\rm e}=38.8-271.8$~cm$^{-3}$ and $n_{\rm
e}=13.5-738.9$~cm$^{-3}$, respectively. Log-normal distributions have
been used to characterize the electron density distribution of
the warm ionized medium \citep{rhc_berkhuijsen08, rhc_hill08,
rhc_redfield08}. 

Then, if we replace $dV$ in Eq.~(\ref{eq:Llambda}) using Eq.~(\ref{eq:ne_dist}) we can express the power radiated in a line as

\beq
\begin{split}
L_\lambda=\frac{Q_{0}}{\alpha_{B}\sqrt{2\pi\sigma^2}} \left[\frac{\rm N^+}{\rm H^+}\right]  \int \exp\left[-\frac{(\ln(n_e/n_{e0}))^2}{2\sigma^2}\right]  \\
\times \frac{P_\lambda(n_e)}{n_{e}}  d\ln n_e.
\end{split}
\eeq

\begin{figure}
\epsscale{1}
\plotone{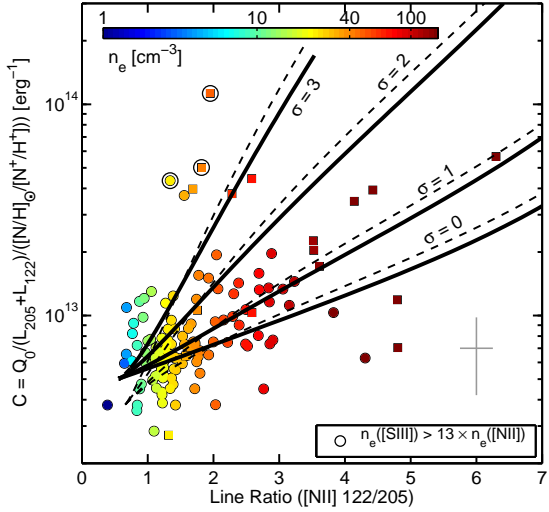}
\caption{Number of ionizing photons per erg of emitted energy in the [NII] lines, $C$, as a function of the \nit~122/205 line ratio, for four values of the electron density distribution parameter: $\sigma=0$ (uniform density), 1, 2 and 3. Central regions with deprojected distances less than 400 pc are shown as squares. The error bars in the bottom corner show the typical uncertainty on the measurements. The solid and dashed lines are based on the \cite{rhc_tayal11} and \cite{rhc_hudson05} collision coefficients, respectively. As the \nit~122/205 line ratio increases, the parameter $C$ becomes increasingly sensitive to the distribution of electron densities. This behavior has a direct effect on the accuracy of the determination of SFRs based on the \nit\ lines, as ${\rm SFR} \propto C \times L_{122+205}$ (see Equation~\ref{eq:SFR}). We also include the measured calibration coefficient $C$ for 141 regions in our sample for which we have \nit~122/205 line ratios available. For each region we use the characteristic oxygen abundance of its parent galaxy as a proxy for the  nitrogen abundance.  Finally, we use open circles to indicate three regions that show high $C$ values and have \siii-based electron density measurements that are a factor of $\sim13$ or higher than those based on the \nit\ lines. This could be an indication that these regions have wide electron density distributions. \label{ne_C}}
\end{figure}

\noindent The observed line ratio is then a function of both $n_{e0}$ and $\sigma$

\beq 
\label{eq:ratio}
\frac{L_{122}}{L_{205}}
=
\frac{\int d\ln n_e \exp[-(\ln(n_e/n_{e0}))^2/2\sigma^2]P_{122}(n_e)/n_e}
     {\int d\ln n_e \exp[-(\ln(n_e/n_{e0}))^2/2\sigma^2]P_{205}(n_e)/n_e}
     ~~~,
\eeq

\noindent and the radiative recombination rate $Q_0$ can be expressed as

\beq
Q_0 = C \left(\frac{({\rm N}/\Ha)_\odot}{{\rm N}^{+}/\Ha^+}\right)
\left[L_{205}+L_{122}\right]
~~~,
\eeq

\noindent where ${\rm N^{+}}/\Ha^+$ is the ionized gas phase abundance ratio and 

\beq
\begin{split}
\label{eq:C}
C(n_{e0},\sigma) = \frac{\alpha_B}{(N/H)_\odot} \times ~~~~~~~~~~~~~~~~~~~~~~~~~~~~~~~~~~~~~~~~~~~~~\\
        \frac{\sqrt{2\pi}\sigma}
        {\int d\ln n_e (P_{205}(n_e)\!+\!P_{122}(n_e))
         \exp[-(\ln(n_e/n_{e0}))^2/2\sigma^2] / n_e}.
\end{split}
\eeq

\noindent Here we introduce the calibration coefficient $C$, which is the number of ionizing photons per erg of emitted energy in the \nit\ lines. $C$ is a function of $n_{e0}$ and the distribution parameter $\sigma$, but it can also be regarded as a function of the observed line ratio $L_{122}/L_{205}$ and $\sigma$, $C(L_{122}/L_{205},\sigma)$. The calculations assume a solar abundance (N/H)$_\odot=7.4\times10^{-5}$ \citep{rhc_asplund09}.

Thus,

\beq \label{eq:SFR}
\begin{split}
\frac{\rm SFR}{{\rm \Msol\yr^{-1}}} = \frac{2.77\times10^{-7}}{f_{\rm ion}}
\left(\frac{C(L_{122}/L_{205},\sigma)}{10^{13}\erg^{-1}}\right) \\
\times \left(\frac{({\rm N/H})_\odot}{\rm N^{+}/H^{+}}\right)
\frac{L_{122}+L_{205}}{\Lsol}
~~~.
\end{split}
\eeq

In summary, we expect that a \nit-based SFR calibration will depend on the nitrogen abundance and the calibration coefficient $C$ (which in turn is a function of the \nit~122/205 line ratio, or $n_{\rm e}$, and the width of the distribution of electron densities).In addition, and similar to what we discussed for the low-density limit case, this calibration will underestimate the star formation rate when the dominant ionization stage is not N$^+$, which is the case for gas ionized by very early-type O stars in high density \hii\ regions.

\begin{figure*}
\epsscale{1}
\plotone{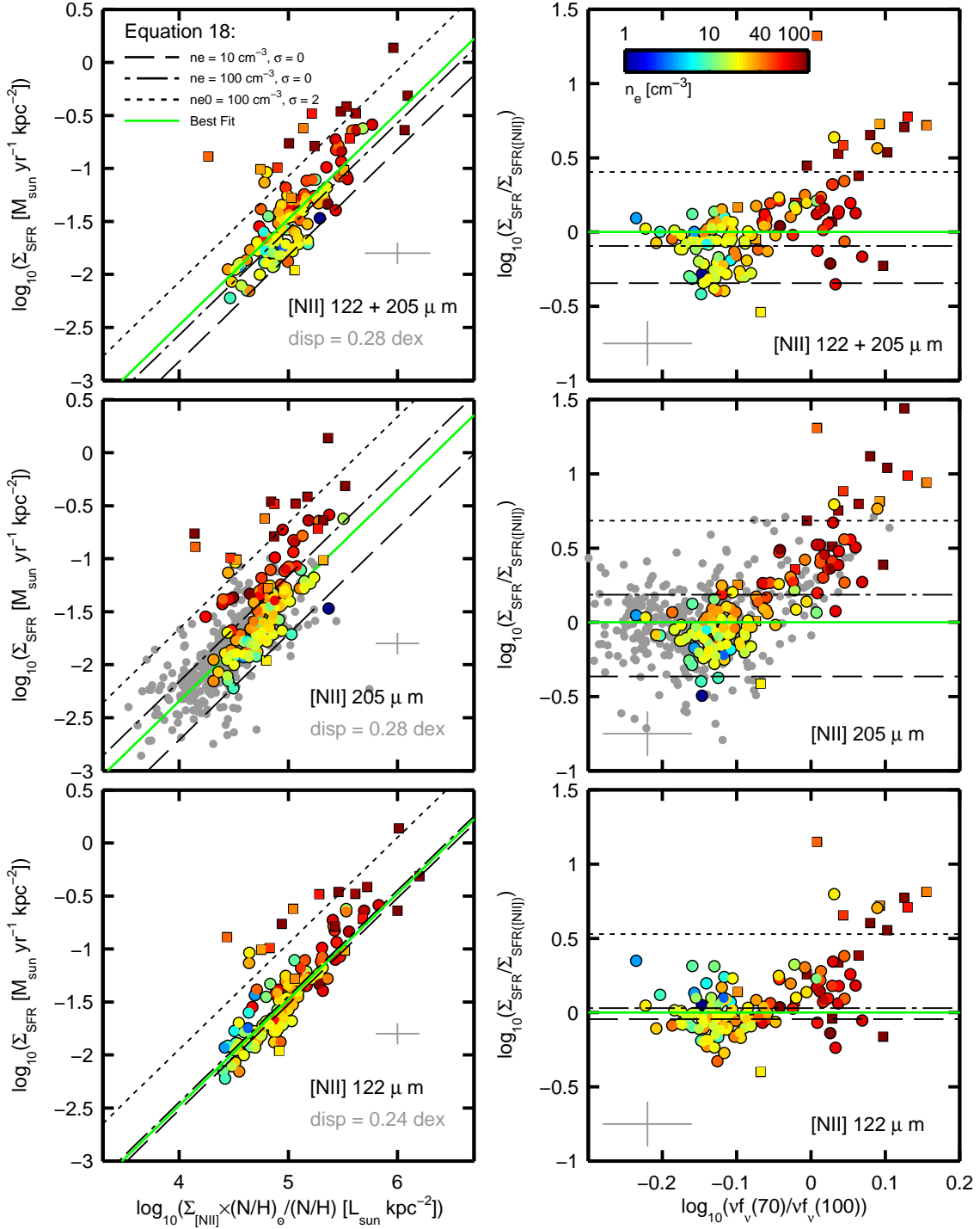}
\caption{(Left panels)  Star formation rate surface density ($\Sigma_{\rm SFR}({\rm H}\alpha+24~\mu {\rm m})$) versus the combined ($\Sigma_{122+205}$) and individual ($\Sigma_{122}$ and $\Sigma_{205}$) \nit\ line surface brightness scaled by the nitrogen abundance factor (N/H)$_{\odot}$/(N/H). Each point represents a $17\arcsec$ region selected from the 21 BtP galaxies.  The error bars in the bottom corner show the typical uncertainty on the measurements. For regions with \nit~122/205 line ratios available, the color represents the electron density. In the case of the \niii\ line, the gray points correspond to 366 additional regions that lie outside the KINGFISH/PACS coverage and therefore lack \nii\ observations. We also show regions with deprojected distances to the center smaller than 400~pc as squares. The green solid line corresponds to the best fit through the data for a fixed slope of one. The rest of the line correspond to results from Equation~(\ref{eq:SFR}) for different assumptions on the electron density distribution parameters $n_{\rm e}$ and $\sigma$. (Right panels) Ratio between $\Sigma_{\rm SFR}$ measured using H$\alpha$+24~$\mu$m and the \nit\ lines (based on the best linear fit) as a function of infrared color and electron density. We also include the predictions from Equation~(\ref{eq:SFR}) shown in the left panels. We observe a systematic increase of the $\Sigma_{\rm SFR}/\Sigma_{\rm SFR}$(\nit) ratio as a function of infrared color and electron density. As predicted by Equation~(\ref{eq:SFR}), these deviations could be associated to a higher density distribution parameter $\sigma$, or to electron densities higher than the critical density of the \nit\ lines. We observe the largest deviations from the linear fit for the \niii\ line, which is expected given that the critical density of this line is about six times lower than the critical density of the \nii\ transition. \label{nii_sfr}}
\end{figure*}

Figure~\ref{ne_C} shows the behavior of $C$ as a function of the \nit~122/205 line ratio and $\sigma$. The solid and dashed lines represent the predictions based on the \cite{rhc_tayal11} and \cite{rhc_hudson05} collision coefficients, respectively. The first thing we notice is that, for a fixed value of $\sigma$, $C$ increases as a function of the \nit~122/205 line ratio (or electron density). This is the direct effect of the collisional suppression of the \nit\ emission at electron densities near and above the critical density of the transition. In this regime, \nit\ collisional de-excitations compete with the radiative decays and the \nit\ intensity starts to systematically underestimate the amount of star formation activity. In practical terms, this means that two regions with the same SFR, similar electron density distribution, but different \nit~122/205 line ratio, then the one with the higher line ratio (or higher $n_{\rm e}$) will have a lower $L_{122+205}$ luminosity.  This is especially true for cases when the electron density is higher than the critical density of the \niii\ line, $n_{\rm e}\gtrsim44$~cm$^{-3}$ (\nit~$122/205\gtrsim1.6$). 

The second thing to note in Figure~\ref{ne_C} is that the calibration coefficient $C$ becomes increasingly sensitive to the distribution of electron densities $\sigma$. For example, for an observed line ratio \nit~$122/205=4$, the value of $C$ increases by a factor $\sim$5 as $\sigma$ varies from 0 to 2. This implies than in galaxies with $L_{122}/L_{205} \gtrsim 2$, the inferred SFR will be uncertain unless there is additional information available (e.g., from observations of other emission line ratios) to constrain the actual distribution of electron densities.

Finally, Figure~\ref{ne_C} also includes the calibration coefficient $C$ measured in regions with \nit~122 and 205~$\mu$m both measured, and $Q_0$ inferred from $\Sigma_{\rm SFR}({\rm H}\alpha+24~\mu{\rm m})$ using above Equation~(\ref{eq:Q2SFR}). For the nitrogen abundance of the gas relative to the Sun, ${\rm (N^{+}/H^{+})/(N/H)_{\odot}}$, we used as a proxy the oxygen abundance of its parent galaxy, and we assume that nitrogen and oxygen abundances scale linearly. This assumption is consistent with the observed scatter in the ${\rm (N/O)}-{\rm (O/H)}$ correlation \citep[e.g,][]{rhc_groves04,rhc_pm09} in the metallicity range of the BtP galaxies ($12+log_{10}{\rm (O/H)}\sim8.6-9$). If we assume instead of the linear scaling an analytic function dependence of the nitrogen abundance with oxygen --like the one used in the MAPPINGS code--, then we observe a larger spread in the $C-$\nit~122/205 line ratio values of our BtP regions.

In general, the calibration coefficient $C$ predictions are consistent with the observations, and as the \nit~122/205 line ratio increases, the scatter in the observations can be explained by different assumptions on the width of the electron density distribution. A handful of exceptions are the group of 7 regions with \nit~122/205 line ratios around $\sim1.5-2.5$ that show significantly higher $C$ values compared to the theoretical expectations, even for the $\sigma=2$ case. To explore if the reason for the high calibration coefficients measured in these regions is associated to high $\sigma$ values, we add the \siii\ lines to our analysis. Since sulfur has a second ionization potential (23.3~eV) higher than that of \nitnp, the sulfur forbidden lines \siii~18.7 and 33.5~$\mu$m probe a higher ionization gas than the \nit\ infrared lines. \siii~18.7/33.5 is sensitive to changes in the electron density in the $n_{\rm e}\sim100-10^{4}$~cm$^{-3}$ range. For three of the seven regions, fluxes for the \siii~18.7 and 33.5~$\mu$m lines are available \citep{rhc_dale09b} using The Infrared Spectrograph \citep[IRS;][]{rhc_houck04} on board {\it Spitzer}. The size of the aperture used to measure the \siii\ line fluxes is $23\arcsec \times 15\arcsec$, roughly similar to the $\sim17\arcsec$ SPIRE-FTS beam at 205~$\mu$m. We find that the \siii-based electron densities of these regions are at least a factor of $\sim13$ higher than those obtained using the \nit\ lines. This suggests that these regions could have a wide distribution of electron densities, where a higher-ionization, higher-density gas component powered by massive stars coexists with the more diffuse gas traced by the \nit\ lines. A complete analysis of the distribution of electron densities based on the combination of multiple tracers of ionized gas density (e.g., \nit, \siii, [\ion{O}{3}]) will be presented in a future paper.

\subsection{Correlations between the \nit~122 and 205~$\mu$m transitions and the star formation activity}

In this section we continue the study of the relationship between \nit\ emission and star formation activity, but this time we also include the individual correlations with the \nit~122 and 205~$\mu$m line surface brightness ($\Sigma_{122}$ and $\Sigma_{205}$, respectively), as for many sources only one of these two lines will be available.

The first panel in Figure~\ref{nii_sfr} shows the correlation between $\Sigma_{\rm SFR}({\rm H\alpha+24~\mu m})$ and $\Sigma_{122+205}$ scaled by the nitrogen abundance factor (N/H)$_{\odot}$/(N/H). The best linear fit yields:

\begin{equation}
\frac{\Sigma_{\rm SFR}}{{\rm M_{\odot}~yr^{-1}}~{\rm kpc}^{-2}}=3.31\times10^{-7} \left(\frac{({\rm N/H})_\odot}{\rm N^{+}/H^{+}}\right)\frac{\Sigma_{122+205}}{L_{\odot}~{\rm kpc}^{-2}}.
\end{equation}

\noindent Based on this fit, the right panel shows the scatter as a function of IR color. We also plot the  relation from Equation~(\ref{eq:SFR}) for three cases: two uniform density cases ($\sigma=0$) with electron densities of $n_{\rm e} =10$ and 100~cm$^{-3}$, and one case with a wider electron density distribution ($\sigma=2$) and $n_{\rm e0}=100$~cm$^{-3}$. For the nitrogen abundance term in Equation~(\ref{eq:SFR}), we use as a proxy the median oxygen abundance of the sample ($12+{\rm log_{10}(O/H)} \approx 8.83$). We find that the best linear fit to the data lies between the expectations from the $n_{\rm e}=10$ and $n_{\rm e}=100$~cm$^{-3}$, single density ($\sigma=0$) models (long-dashed and dot-dashed lines).
We also observe that the $\Sigma_{\rm SFR}/\Sigma_{\rm 122+205}$ ratio tends to increase as a function of infrared color and electron density. As we discussed in Section~3.5, these deviations can be understood in terms of the electron density of the gas relative to the critical density of the \nit\ transitions, and the density distribution parameter $\sigma$. For example, from Equation~(\ref{eq:SFR}) we expect regions with a density distribution parameter $\sigma=2$ and $n_{\rm e0}=100$~cm$^{-3}$ (short-dashed line) to have $\Sigma_{\rm SFR}/\Sigma_{\rm 122+205}$ ratios a factor of $\sim5$ higher than the ratio found by the best linear fit. In our sample, there are two galaxy central regions with densities around $n_{\rm e}\approx100$~cm$^{-3}$ that have $\Sigma_{\rm SFR}/\Sigma_{\rm 122+205}$ ratios that are consistent with the $\sigma=2$, $n_{\rm e0}=100$~cm$^{-3}$ model. 

In the case of the \niii\ emission, we include in our analysis 366 additional regions that lie outside the KINGFISH/PACS coverage (meaning, they lack \nii\ observations -- see middle row of Figure~\ref{nii_sfr}). The best linear fit through the $\Sigma_{205}-\Sigma_{\rm SFR}$ correlation yields:

\begin{equation}
\frac{\Sigma_{\rm SFR}}{{\rm M_{\odot}~yr^{-1}}~{\rm kpc}^{-2}}=4.51\times10^{-7} \left(\frac{({\rm N/H})_\odot}{\rm N^{+}/H^{+}}\right)\frac{\Sigma_{205}}{L_{\odot}~{\rm kpc}^{-2}}.
\end{equation}

\noindent The scatter around this fit is 0.28~dex, similar to the scatter in the $\Sigma_{122+205}-\Sigma_{\rm SFR}$ correlation. 
We also observe a larger spread in the results from Equation~(\ref{eq:SFR}) for the same set of assumptions on $n_{\rm e}$ and $\sigma$ as in the case of the combined $\Sigma_{122+205}$ emission.
For regions with infrared colors cooler than $\nu f_{\nu}(70~\mu{\rm m})/\nu f_{\nu}(100~\mu{\rm m})\approx0.9$ the scatter cloud around the linear fit is roughly symmetric, with a standard deviation of 0.25~dex.  At about the same infrared color threshold, \cite{rhc_rhc15} find a systematic increase of the fit residuals from the $\Sigma_{\rm [CII]}-\Sigma_{\rm SFR}$ correlation, and \cite{rhc_croxall12} find a drop in the \cii\ to FIR ratio for regions in NGC 4559 and NGC 1097 (although they do not find any signs of a ``\nit-deficit''). 
For IR colors warmer than $\nu f_{\nu}(70~\mu{\rm m})/\nu f_{\nu}(100~\mu{\rm m})\approx0.9$ --which according to Equation~(\ref{eq:correlation_uncorr}) corresponds to an electron density of $n_{\rm e}\approx33$~cm$^{-3}$, close to the critical density of the line-- we observe a strong increase in the $\Sigma_{\rm SFR}/\Sigma_{\rm 205}$ ratio as a function of infrared color and electron density. These deviations are mainly driven by the collisional quenching of the \niii\ line due to its relatively low critical density.

Finally, in the case of the $\Sigma_{122}-\Sigma_{\rm SFR}$ correlation, shown in the bottom two panels of Figure~\ref{nii_sfr}, the best fit through the data using a fixed slope of one yields:

\begin{equation}
\frac{\Sigma_{\rm SFR}}{{\rm M_{\odot}~yr^{-1}}~{\rm kpc}^{-2}}=3.33
\times10^{-7} \left(\frac{({\rm N/H})_\odot}{\rm N^{+}/H^{+}}\right)\frac{\Sigma_{122}}{L_{\odot}~{\rm kpc}^{-2}}.
\end{equation}

\noindent This linear fit is similar to the relation from Equation~(\ref{eq:SFR}) for both single density, $n_{\rm e}=10$ and 100~cm$^{-3}$ models. The scatter around the fit is 0.24~dex, slightly better than the 0.28~dex we measure in the $\Sigma_{205}-\Sigma_{\rm SFR}$ correlation. Compared to the \niii\ case, regions start to deviate from the best linear fit at warmer IR colors, and there seems to be a smaller dependence between the amplitude of these deviations and the electron density of the region. This is expected as the critical density of the \nii\ transition is about six times higher than that of the \niii\ line. However, and based on the relatively low critical densities of some of the regions that deviate $\gtrsim0.5$~dex compared to the \nii\ critical density, it seems that the density effect by itself is not sufficient to explain these deviations. As we discussed in \S4.2, additional physical effects that could be playing a role in this warm regions are the hardness of the radiation field (that sets the ${\rm N}^{++}/{\rm N}^{+}$ ratio), and the width of the electron density distribution.

\begin{figure*}
\epsscale{1.1}
\plotone{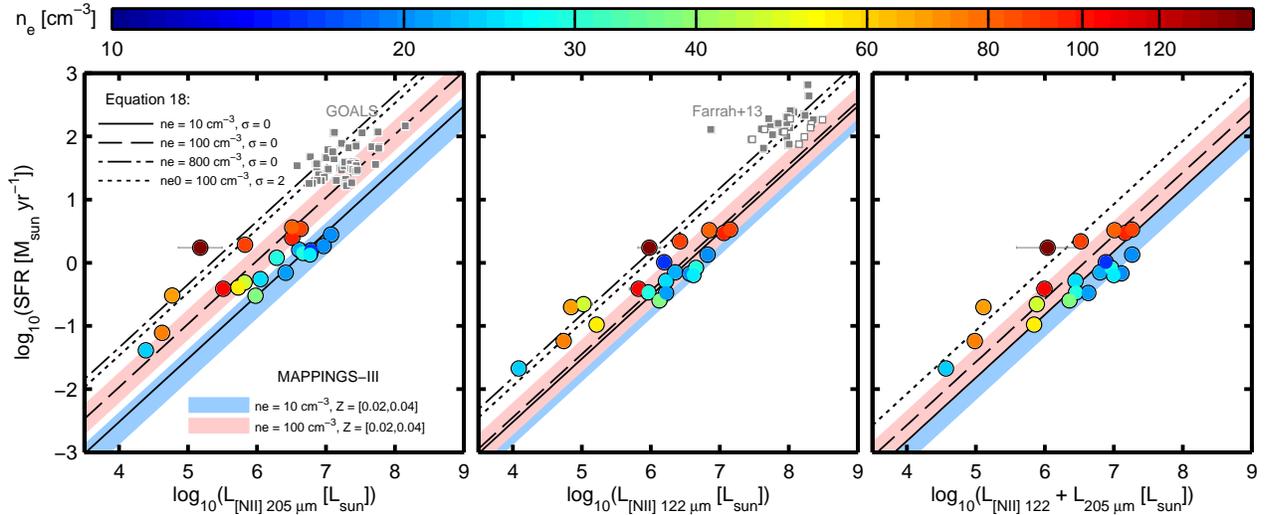}
\caption{\nit~122 and 205~$\mu$m luminosities versus SFR compared to the results from the MAPPINGS-III code and Equation~(\ref{eq:SFR}). BtP galaxies are shown as circles, where the color represents the electron density. In general, we find a good agreement between the observed correlations and the model predictions, where the scatter of the correlation can be understood in terms of variations in properties of the ionized gas ($n_{\rm e}$ and $\sigma$) and metallicity. {\it (Left)} SFR versus $L_{205}$. 
For comparison, we include LIRGs from the GOALS sample observed in \niii\ emission using {\it Herschel}/SPIRE-FTS \citep[][]{rhc_zhao13} .
The colored areas show the MAPPINGS-III results based on assuming \hii\ regions with electron densities of $n_{\rm e}=10$ (blue) and 100~cm$^{-3}$ (red) and gas metallicities in the $Z=Z_{\odot}-2Z_{\odot}$ range. Finally, the different black lines show the results from Equation~(\ref{eq:SFR}) for the following set of conditions: $\sigma=0$ and $n_{\rm e}=10$~cm$^{-3}$ (solid line), $\sigma=0$ and $n_{\rm e}=100$~cm$^{-3}$ (dashed line), $\sigma=0$ and $n_{\rm e}=800$~cm$^{-3}$ (dot-dashed line), and $\sigma=2$ and $n_{\rm e0}=100$~cm$^{-3}$ (short-dashed line). {\it (Center)} SFR versus $L_{122}$. We include for comparison the \cite{rhc_farrah13} sample of ULIRGs observed in \nii\ emission using {\it Herschel}/PACS. Open squares correspond to 3-$\sigma$ upper limits in $L_{122}$. The model and Equation~(\ref{eq:SFR}) results are based on the same set of assumptions than those adopted in the first panel. {\it (Right)} Similar to the first two panels, but this time we plot the SFR versus the combined \nit\ luminosity, $L_{122+205}$. 
\label{niisfr_comp}}
\end{figure*}

The \cii~158~$\mu$m transition is another far-infrared line proposed to study the star formation activity in local and high-$z$ galaxies \citep[e.g.,][]{rhc_rhc15}. The scatter in the $\Sigma_{\rm [CII]}-\Sigma_{\rm SFR}$ correlation is about $\sim0.2-0.3$~dex \citep[][]{rhc_delooze14,rhc_rhc15}, similar to the scatter measured in the $\Sigma_{\rm [NII]}-\Sigma_{\rm SFR}$ correlations that include the \nii\ transition.
The key physical advantage of the \cii\ transition is that it is the dominant coolant of the neutral atomic and molecular phases \citep[e.g.,][]{rhc_rosenberg15} so, unlike the \nit, its intensity
reflects the heating of the star forming gas. As long as the majority of the \cii\ emission arises from neutral atomic and molecular gas, and wherever heating of these phases is dominated by star formation activity, the \cii\ intensity provides a direct measure of that activity modulo an uncertain heating efficiency and the cooling contribution from other potentially important coolants such as \oi~63~$\mu$m emission.

\subsection{Comparison with models and other extragalactic \nit\ samples}

In this section we continue the study of the relation between \nit\ emission, star formation activity and electron density by complementing our sample of nearby spirals with local LIRGs, and comparing these observations to predictions from Equation~(\ref{eq:SFR}) and the photoionization code MAPPINGS-III \citep{rhc_levesque10}. A brief description of the LIRG sample and the code MAPPINGS-III can be found in Section~2.

The first panel in Figure~\ref{niisfr_comp} shows the $L_{205}-{\rm SFR}$ correlation for the BtP galaxies and local LIRGs drawn from the GOALS \citep{rhc_zhao13} sample. The luminosities and SFRs for the BtP galaxies are the result of the sum of the individual regions shown in Figure~\ref{nii_sfr}; the color of the circles indicate the average electron density of the galaxy. 
We observe that the average electron density of BtP galaxies range between $n_{\rm e}=15$ and 100~cm$^{-3}$, and those systems with electron densities higher than the critical density of the line tend to have higher SFR/$L_{205}$ ratios. This result is similar to what we find in the analysis of the spatially resolved $\Sigma_{205}-\Sigma_{\rm SFR}$ correlation. 

In addition to the observations, Figure~\ref{niisfr_comp} includes results from Equation~(\ref{eq:SFR}) and the MAPPINGS-III code. For the latter, the shaded color areas represent the model results for different assumptions on the electron density (blue for $n_{\rm e}=10$~cm$^{-3}$ and red for $n_{\rm e}=100$~cm$^{-3}$) and metallicity of the gas (lower and upper boundaries for $Z=2Z_{\odot}$ and $Z=Z_{\odot}$, respectively). The MAPPINGS-III code predicts that, for a fixed metallicity, \hii\ regions with electron densities of $n_{e}=100$~cm$^{-3}$  will have ${\rm SFR}/L_{205}$ ratios a factor of $\sim4$ higher than those with $n_{e}=10$~cm$^{-3}$. These results are consistent with the observed $L_{205}-{\rm SFR}$ relationship for the BtP galaxies. As we discussed in Section 3.5, one of the reasons for this behavior is the collisional quenching of the \niii\ line when $n_{e} \gtrsim n_{\rm crit}$. 

Regarding the GOALS galaxies, about half of them have ${\rm SFR}/L_{205}$ ratios consistent with the MAPPINGS-III results for \hii\ regions with $n_{e}=100$~cm$^{-3}$. The other half have ${\rm SFR}/L_{205}$ ratios too high to be interpreted by the MAPPINGS-III model outputs. These systems also tend to have high IRAS $f_{\nu}(60)/f_{\nu}(100)$ colors \citep{rhc_zhao16}. One possibility is that the ionized gas for these cases is denser than $n_{e}=100$~cm$^{-3}$. We explore this scenario using Equation~(\ref{eq:SFR}) and we found that, under the assumption of isodensity gas ($\sigma=0$), an electron density of 800~cm$^{-3}$ is required to reproduce the high ${\rm SFR}/L_{205}$ ratios (dot-dashed line). If we drop the assumption of single density gas, then the high ${\rm SFR}/L_{205}$ ratios can be described by assuming a characteristic electron density of $n_{e0}=100$~cm$^{-3}$ and increasing the density distribution parameter from $\sigma=0$ to $\sigma=2$. In addition to the density effect, another factor that could play a role is the hardness of the ionizing radiation field, which controls the N$^{++}$/N$^{+}$ ratio. Evidence for this comes from the observed systematic decrease in the $L_{205}/L_{\rm IR}$ ratio of star-forming galaxies and (U)LIRGs as the $L_{\rm [OIII]}/L_{\rm 205}$ ratio increases \citep{rhc_zhao13}. The $L_{\rm [OIII]}/L_{\rm 205}$ ratio should resemble the N$^{++}$/N$^{+}$ ratio, as the energy needed to form N$^{++}$ is only $\sim6$~eV higher than that needed to form O$^{++}$. 
This analysis reinforces our previous conclusion about the limitations on using the \niii\ line as a star formation tracer when the properties of the ionized gas of the source (in particular $n_{\rm e}$, $\sigma$, the ionization parameter and the hardness of the radiation field) are not constrained.

Compared to the \niii\ case, the observed correlations for the BtP galaxies that involve the \nii\ line are tighter, resulting from a weaker dependence of the \nii\ line emission with electron density. This result is consistent with the predictions from the MAPPINGS-III code, where, for example, the ${\rm SFR}/L_{122}$ ratios for the $n_{e}=10$ and 100~cm$^{-3}$ cases only differ by a factor of 1.5. Similarly, the results from Equation~(\ref{eq:SFR}) for $\sigma=0$ and $n_{e}=10$ and 100~cm$^{-3}$ differ only by a factor of 1.2. Regarding the ULIRGs drawn from the \cite{rhc_farrah13} sample, 
they tend to have ${\rm SFR}/L_{122}$ ratios that are a factor of $\sim3$ higher than those of the BtP galaxies. According to the results from Equation~(\ref{eq:SFR}), this difference between the BtP galaxies and the LIRGs suggests that the ionized gas in the latter is dominated by dissimilar conditions, e.g., an homogeneous ionized medium with a high electron density ($n_{e}=800$~cm$^{-3}$), or ionized gas with $n_{e0}=100$~cm$^{-3}$ but characterized by a wide electron density distribution ($\sigma=2$). Similar to the case of the (U)LIRGs in the \cite{rhc_zhao13} sample, one additional factor that needs to be considered is the increase in the N$^{++}$/N$^{+}$ ratio with the hardness of the ionizing radiation field. In particular,  Seyfert 2 and ULIRGS in the \cite{rhc_farrah13} sample that show weak Polycyclic Aromatic Hydrocarbon (PAH) emission (probably a sign of intense radiation fields) likely have hard radiations fields and $G_0/n_e \approx 10~{\rm cm}^3$. These conditions place those galaxies in the regime where N$^{++}$ is the dominant ionization stage \citep{rhc_abel09}, which may explain the observed high ${\rm SFR}/L_{122}$ ratios compared to those measured in the BtP sample.

In summary, we find that a \nit-based SFR calibration (Equation~\ref{eq:SFR}) depends, among other things, on the electron density of the gas, the shape of the electron density distribution, hardness of the radiation field, and the nitrogen abundance. Additional constraints on the properties of the ionized gas, provided for example by the combination of the \nit\ and \siii\ lines, are required to avoid uncertainties in the SFR determinations larger than a factor of $\sim2$. Without prior knowledge on these variables, the \nii\ transition is a slightly more reliable tracer than the \niii\ line due to its approximately six times higher critical density.

\section{{\bf Summary and Conclusions}} 

In this paper we use the \nit~122 and 205~$\mu$m far-infrared transitions to study the properties of the low-excitation \hii\ gas in 21 galaxies observed by {\it Herschel} as part of the ``Beyond the Peak'' and KINGFISH projects. In particular, we use the \nit~122/205 line ratio to measure the electron density of the low-excitation \hii\ gas. We then study the dependence between the electron density and properties of the ISM such as radiation field strength, infrared color, metallicity, among others. The \nit~122 and 205~$\mu$m far-infrared lines also have potential as star formation tracers, and in this work we study the correlations between the \nit~122 and 205~$\mu$m emission and the star formation activity. In particular, we explore the dependence of these correlations with properties of the ionized gas such as its density and metallicity.

We highlight the following points: 

\begin{enumerate}
\item For 141 regions selected from 21 galaxies we measure \nit~122/205 line ratios in the range $\sim0.6-6$, which correspond to electron densities of the photoionized gas in the range $n_{\rm e}\sim3-300$~cm$^{-3}$ (assuming a single $n_{\rm e}$ within each region). If we consider instead a distribution of electron densities that follows a log-normal distribution characterized by a width $\sigma$, and a characteristic electron density $n_{\rm e0}$ (Equation~\ref{eq:ne_dist}), then the relationship between the \nit~122/205 line ratio and $n_{\rm e0}$ is given by Equation~\ref{eq:ratio}. We find that only one region has a \nit~122/205 line ratio below the theoretical limit of $\sim0.6$, which corresponds to gas with $n_{\rm e} \ll n_{\rm crit}$. The median ionized gas density in the sample is $n_{\rm e}\approx30$~cm$^{-3}$, comparable to the median electron density measured in the Carina nebula \citep{rhc_oberst06,rhc_oberst11}. Within individual galaxies we measure variations in the ionized gas density as high as a factor of $\sim50$. In general, the central $\sim400$~pc regions exhibit the highest electron densities in the galaxy, which is expected as they typically have the highest star formation rate surface densities. 

\item We find a good correlation between electron density and infrared color ($\nu f_{\nu}(70~\mu{\rm m})/\nu f_{\nu}(100~\mu{\rm m})$), dust-weighted mean starlight intensity ($\langle U \rangle$), TIR surface density ($\Sigma_{\rm TIR}$) and SFR surface density ($\Sigma_{\rm SFR}$). The origin of these good correlations could be associated with: (1) the evolutionary stage of the \hii\ region, as young, compact \hii\ regions will produce very intense radiation fields, or (2) the fact that dense \hii\ regions in pressure equilibrium with the surrounding neutral gas implies a high density molecular gas environment that may lead to further star formation activity. 

These relationships can also be useful when, in order to predict the \niii\ intensity from the \nii\ intensity (or vice versa), an electron density needs to be assumed. In particular, we provide an equation to estimate the electron density from the $\nu f_{\nu}(70~\mu{\rm m})/\nu f_{\nu}(100~\mu{\rm m})$ infrared color:

\begin{align*}
{\rm log_{10}} \left(\frac{n_{e}}{{\rm cm^{-3}}}\right) = 3.41 \times {\rm log_{10}}\left(\frac{\nu f_{\nu}(70)}{\nu f_{\nu}(100)}\right)+1.68.
\end{align*}

\item We use the \nit\ far-infrared transitions to measure the ionizing photon rate $Q_{0}$. In the low-density limit (Equation~\ref{eq:Q_low}) we use the observed \niii\ observations to estimate a median global ionizing photon rate $Q_0=1.98\times10^{52}~{\rm s}^{-1}$ for the BtP galaxies. Because collisional deexcitation has been neglected, this is only a lower bound on $Q_0$. Normalized by the covered area, the median ionizing photon rate surface density is  $1.96\times10^{51}$~s$^{-1}$~kpc$^{-2}$, which is $\sim3$ times the ionizing photon rate measured inside the solar circle ($\sim8.5$~kpc) in the Milky Way \citep{rhc_bennett94}.

\item We derive relations between the \nit\ line emission and star formation rate in the low density limit (Equations~\ref{eq:SFR_low1} and \ref{eq:SFR_low2}), and for distributions of electron densities (Equation~\ref{eq:C} and \ref{eq:SFR}). The latter assumes a log-normal distribution of electron densities characterized by the width of the distribution $\sigma$ ($\sigma=0$ corresponds to uniform density), and the characteristic electron density of the ionized gas $n_{\rm e0}$. We then relate the SFR and the \nit\ luminosity via the the calibration coefficient $C$ (Equation~\ref{eq:C}), which is the number of ionizing photons per erg of emitted energy in the \nit\ lines. $C$ is a function of the \nit~122/205 line ratio (or electron density) and $\sigma$. We find that $C$ increases as a function of both the \nit~122/205 line ratio and the density distribution parameter $\sigma$. The differences between the values of $C$ for different ionized gas conditions can be significant, and imply that not only the \nit~122/205 line ratio, but additional constraints on the electron density distribution are important to accurately measure SFRs based on the \nit\ transitions.

\item In general, we find good correlations between the surface brightness of the \nit\ line emission and the star formation rate surface density. However, and as we show in Equation~(\ref{eq:SFR}), these correlations depend on the nitrogen abundance, the characteristic electron gas density and the density distribution parameter $\sigma$. The lack of constraints on any of these parameters can result in large uncertainties in the SFR determination based on the \nit\ lines only. The best linear fit to the observed correlations of $\Sigma_{\rm SFR}$ with $\Sigma_{\rm 205}$, $\Sigma_{122}$ and $\Sigma_{122+205}$ for the BtP regions are:

\begin{align*}
\frac{\Sigma_{\rm SFR}}{{\rm M_{\odot}~yr^{-1}}~{\rm kpc}^{-2}}
=4.51\times10^{-7} \left(\frac{({\rm N/H})_\odot}{\rm N^{+}/H^{+}}\right) \frac{\Sigma_{205}}{L_{\odot}~{\rm kpc}^{-2}}, \\
=3.33\times10^{-7} \left(\frac{({\rm N/H})_\odot}{\rm N^{+}/H^{+}}\right) \frac{\Sigma_{122}}{L_{\odot}~{\rm kpc}^{-2}}, \\
=3.31\times10^{-7} \left(\frac{({\rm N/H})_\odot}{\rm N^{+}/H^{+}}\right) \frac{\Sigma_{122+205}}{L_{\odot}~{\rm kpc}^{-2}}.
\end{align*}

For all three correlations we find that regions with warm infrared colors tend to show deviations from the best linear fit in the sense that a \nit-based SFR calibration will underestimate the reference amount of star formation activity (measured as a combination of 24~$\mu$m and H$\alpha$ emission). For the \niii\ line, these deviations starts at an IR color threshold of $\nu f_{\nu}(70~\mu{\rm m})/\nu f_{\nu}(100~\mu{\rm m})\approx0.9$, and they increase as a function of $\nu f_{\nu}(70~\mu{\rm m})/\nu f_{\nu}(100~\mu{\rm m})$ and $n_{\rm e}$ until reaching deviations of $\sim1$~dex at $\nu f_{\nu}(70~\mu{\rm m})/\nu f_{\nu}(100~\mu{\rm m})\approx1.2$. This is primarily a density effect, given that in regions with gas densities closer or greater than $n_{\rm crit}$, the \niii\ collisional de-excitations compete with the radiative decays and the \niii\ intensity stops tracing any increment in the star formation activity. However, it is probably in part also due to ionization of N$^{+}$ to N$^{++}$ and N$^{+++}$ in high-density \hii\ regions ionized by early-type O stars.

For the two correlations that involve the \nii\ line we also observe deviations from the best linear fit growing with IR color, but at a higher IR color threshold than in the \niii\ case ($\nu f_{\nu}(70~\mu{\rm m})/\nu f_{\nu}(100~\mu{\rm m}\approx1.1$). This is because the critical density of the \nii\ transition is $\sim6$ times higher than that of \niii, which makes the \nii\ transition less sensitive to the effects of density than the \niii\ line.

\item We compare the $L_{\rm [NII]}-{\rm SFR}$ correlations for the BtP galaxies to a sample of local (U)LIRGs \citep{rhc_zhao13,rhc_farrah13} and predictions from the MAPPINGS-III photoionization code and Equation~(\ref{eq:SFR}) for different assumptions on the ionized gas properties. In general, we find good agreement between the $L_{\rm [NII]}-{\rm SFR}$ correlations and the model results, where the observed trends and scatter can be understood in terms of variations of the electron density of the gas and the width of the electron density distribution. Both theory and observations reveal the importance of having prior knowledge of the ionized gas properties of the source (e.g., metallicity, \nit\ and \siii\ based electron density measurements, ionization parameter) in order to avoid underestimating the star formation activity. In case there are no constraints on the ionized gas, the \nii\ emission could be a slightly more reliable tracer than \niii\ due to its higher critical density. 

The study of the \nit-based SFR calibrations presented in this paper may prove useful to characterize high-redshift ($z\gtrsim2$) galaxies observed by ALMA. As we discussed in \S5, this should be done carefully, as the \nit-based SFR calibration depends on the nitrogen abundance, the electron density, the shape of the electron density distribution, and the ionization state of the gas. In the case there are additional tracers that can be used to measure the SFR (e.g., $L_{\rm FIR}$), the comparison to the \nit-based SFR measurement could be used to constrain the properties of the ionized gas of the system.

\end{enumerate}

\bigskip

We thank the anonymous referee for helpful suggestions that improved the paper. R.H.C. acknowledges support from a Fulbright-CONICYT grant. A.D.B. acknowledges partial support from a CAREER grant NSF-AST0955836, from NASA-JPL 1373858, NSF-AST 1412419 and from a Research Corporation for Science Advancement Cottrell Scholar award. Beyond the Peak research has been supported by a NASA/JPL grant (RSA 1427378).  JDS gratefully acknowledges visiting support from the Alexander von Humboldt Foundation and the Max Planck Institute f{\"u}r Astronomie. FST acknowledges financial support from the Spanish Ministry of
Economy and Competitiveness (MINECO) under grant number AYA2013-41243-P. PACS has been developed by a consortium of institutes led by MPE (Germany) and including UVIE (Austria); KU Leuven, CSL, IMEC (Belgium); CEA, LAM (France); MPIA (Germany); INAF-IFSI/OAA/OAP/OAT, LENS, SISSA (Italy); IAC (Spain). This development has been supported by the funding agencies BMVIT (Austria), ESA-PRODEX (Belgium), CEA/CNES (France), DLR (Germany), ASI/INAF (Italy), and CICYT/MCYT (Spain). HIPE is a joint development by the Herschel Science Ground Segment Consortium, consisting of ESA, the NASA Herschel Science Center, and the HIFI, PACS, and SPIRE consortia. SPIRE has been developed by a consortium of institutes led by Cardiff University (UK) and including Univ. Lethbridge (Canada); NAOC (China); CEA, LAM (France); IFSI, Univ. Padua (Italy); IAC (Spain); Stockholm Observatory (Sweden); Imperial College London, RAL, UCL-MSSL, UKATC, Univ. Sussex (UK); and Caltech, JPL, NHSC, Univ. Colorado (USA). This development has been supported by national funding agencies: CSA (Canada); NAOC (China); CEA, CNES, CNRS (France); ASI (Italy); MCINN (Spain); SNSB (Sweden); STFC, UKSA (UK); and NASA (USA). This work is based (in part) on observations made with Herschel, a European Space Agency Corner- stone Mission with significant participation by NASA. This research has made use of the NASA/IPAC Extragalactic Database (NED), which is operated by the Jet Propulsion Laboratory, California Institute of Technology, under contract with the National Aeronautics and Space Administration.

\bibliography{/Users/rhc/Documents/references.bib}

\end{document}